\documentclass[lettersize,journal]{IEEEtran}
\usepackage{amsmath,amsfonts,amssymb}
\usepackage{array}
\usepackage[caption=false,font=normalsize,labelfont=sf,textfont=sf]{subfig}
\usepackage{textcomp}
\usepackage{stfloats}
\usepackage{url}
\usepackage{verbatim}
\usepackage{graphicx}
\usepackage{cite}
\usepackage{algpseudocode}
\usepackage{algorithm}
\usepackage{booktabs,physics}
\usepackage{tabularx,tablefootnote,wrapfig,multirow}
\usepackage{hyphenat,xspace,color,colortbl,enumitem}

\newcolumntype{Y}{>{\centering\arraybackslash}X}
\newcommand{\parahead}[1]{\vspace{12pt plus 0pt minus 2pt}\noindent{\bfseries #1}}

\def\R{\mathbb R}

\def\argmax#1{\underset{#1}{\operatorname{arg\,max}}}

\begin{document}

\title{Physics-Inspired Discrete-Phase Optimization for 3D Beamforming with PIN-Diode Extra-Large Antenna Arrays}

% \title{Quantum-Inspired 3D Beamforming with Inexpensive  PIN-Diode Extra-Large Antenna Arrays}

\author{Minsung Kim, \IEEEmembership{Student Member,~IEEE}, Annalise Stockley, Keith Briggs, Kyle Jamieson, \IEEEmembership{Senior Member,~IEEE}
        % <-this % stops a space
\thanks{Minsung Kim and Kyle Jamieson are with the Department of
Computer Science, Princeton University. Annalise Stockley is with the AI Centre, BT, London, and Keith Briggs is with BT Research Labs, Martlesham Heath, UK.}% <-this % stops a space
}



\maketitle

\begin{abstract}

Large antenna arrays can steer narrow beams towards a target area, and thus improve the communications capacity of wireless channels and the fidelity of radio sensing. Hardware that is capable of continuously-variable phase shifts is expensive, presenting scaling challenges. PIN diodes that apply only discrete phase shifts are promising and cost-effective; however, unlike continuous phase shifters, finding the best phase configuration across elements is an NP-hard optimization problem. Thus, the complexity of optimization becomes a new bottleneck for large-antenna arrays. To address this challenge, this paper suggests a procedure for converting the optimization objective function from a ratio of quadratic functions to a sequence of more easily solvable quadratic unconstrained binary optimization (QUBO) sub-problems. This conversion is an exact equivalence, and the resulting QUBO forms are standard input formats for various physics-inspired optimization methods.
% , and the subproblems are solvable by physics-inspired optimization methods. 
We demonstrate that a simulated annealing approach is very effective for solving these sub-problems, and we give performance metrics for several large array types optimized by this technique. Through numerical experiments, we report 3D beamforming performance for extra-large arrays with up to 10,000 elements.
% }
\end{abstract}

\begin{IEEEkeywords}
Wireless Power Transmission, Beamforming, PIN-Diode Array, Discrete Phase, QUBO, Simulated Annealing.
\end{IEEEkeywords}

\section{Introduction}
\label{s:intro}
\IEEEPARstart{B}{eamforming} through an antenna array is a fundamental building block of radio communications and sensing to increase range and to focus transmitted signals to desired locations with minimum undesirable beam interference~\cite{godara1997application,rashid1998joint}.
% in wireless communications to increase the communication range and focus transmitted signals to desired locations with minimum undesirable beam interference, instead of having them spread out in all directions
It is central to systems with antenna arrays, including multi-antenna systems in cellular networks, wireless local area
networks, wireless power transmission (WPT), and radar. As demonstrated in \cite{love2003grassmannian,sohrabi2016hybrid,alrabadi2013beamforming}, beamforming can be far more accurate with larger antenna arrays.
% Many radio systems beamforming in wireless applications can also be used to increase data stream capacity between transmitter and receiver.
The computation of the far-field wireless power generated by an antenna array of radiating elements is standard~\cite{VT4}. For beamforming, it is necessary to define an optimization problem with an appropriate objective function related to the power at the target area~\cite{brown1984history}. A good option is to define a small connected region $R$ on the surface of a large sphere surrounding the array and try to maximize the ratio of energy flux through $R$ to energy flux over the whole sphere $S$~\cite{Oliveri,van2002optimum}. This paper uses such an objective function throughout.

% , which was first introduced in \cite{Oliveri,van2002optimum}.

% Note that this does not explicitly constrain the undesired side-lobes. 
% The formulation was introduced by Oliveri \emph{et al.} \cite{Oliveri}.
% , but it was limited to a half-space and vertical projection of a planar square patch on the sphere. In our work, we extend it fully into three dimensions and also choose $R$ to be a circular patch to consider more realistic scenarios.  

\begin{figure}
    \centering
    \includegraphics[width=0.7\linewidth]{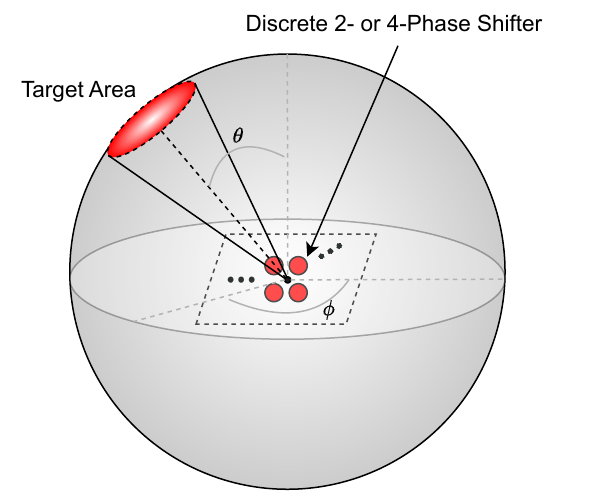}
    \caption{Beamforming with PIN-diode-based low-resolution discrete (2- or 4-) phase shifters, aiming to maximize the power radiated on the target area out of the total transmitted power.} 
\label{f:3d_baemforming}
\end{figure}

This paper addresses methods of solution of transmitter array phase-configuration optimization problems. 
Figure~\ref{f:3d_baemforming} shows our schematic structure. In planar arrays, each radiating antenna element controls a \emph{weight} that multiplies the signal being radiated from the element. This weight encodes both the transmission phase of the element and the power with which it transmits.
% While the original formulation was limited to a half-space and vertical projection of a planar square patch on the sphere, we extend it fully into three dimensions and also choose $R$ to be a circular patch to consider more realistic scenarios. 
The goal of beamforming is to find the optimal weight configuration (\emph{i.e.,} the best combination for all the elements) to beamform more accurately toward a target direction and area. The ideal element permits continuous (analog) weight setting, where every element can transmit with any phase and amplitude, as it optimizes beam gain~\cite{Oliveri}. Furthermore, good methods to solve optimization problems of this type already exist. 
% In other words, the optimization problem for beamforming is simple with continuous weights.
However, engineering an antenna array hardware capable of continuous-phase transmission is challenging and costly, making the size of the array and thus beamforming performance limited. 
% Therefore, we want to enable large-scale antenna array, this

If we consider instead the \emph{discrete phase} case, where each element can transmit with one of a number of predetermined phases (\emph{i.e.,} discrete phase shifts), there is suitable hardware in development that uses an array of inexpensive meta-material-based positive-intrinsic-negative (PIN) diodes to transmit beams, particularly for the 2- or 4-phase case~\cite{wu2019beamforming,huang2017dynamical}. With this hardware, massive antenna arrays can be practically designed.
% meta-material-based wave propagation technique that is realized by manipulating electromagnetic waves has drawn considerable attention for its efficient hardware implementations
% The advent of this hardware has motivated numerous studies on meta-material-based reflective intelligent surfaces (RIS) between transmitter and receiver. With 
However, with limited phases available, the beamforming optimization problem becomes much more mathematically complex to solve.
% , since it is a combinatorial optimization problem. 
As the size of the array increases, or as the number of available discrete phases increases, the complexity of the problem grows exponentially, and we in fact have an NP-hard combinatorial optimization problem. Thus an exhaustive search for the optimal solution is not tractable; for an antenna array with $N^2$ elements, each of which applies $K$ possible phase shifts, $K^{N^2}$ possible phase configurations need to be tested. If solved by
integer linear programming, the maximum practicable number of array elements is about 40, even in the
2-phase case~\cite{stockley2023optimizing}. 
% Thus, for scalable WPT system with discrete phase shifting,  
Especially for mmWave scenarios, a large number of antenna elements are typically expected to steer the beam precisely due to the propagation characteristics of the used bands, making the phase-shifting optimization problem more computationally demanding. 
Thus, computation capabilities become a bottleneck to scalable antenna-array systems with discrete phase shifting.

% that is inspired by Physics principles, which has recently been identified as a promising computing tool . 

\begin{figure}
    \centering
    \includegraphics[width=0.95\linewidth]{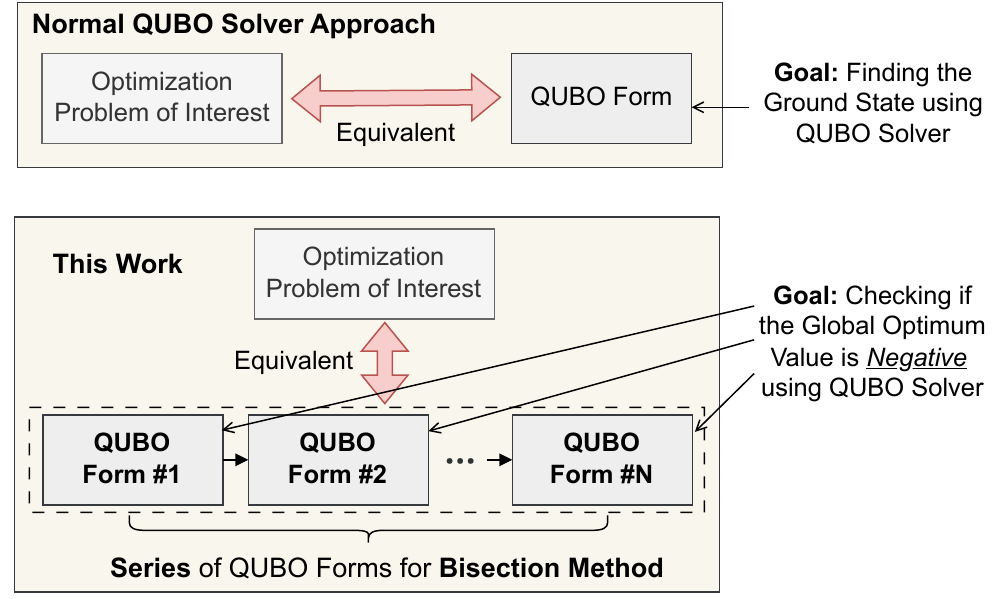}
    \caption{Comparison of the QUBO structure and the QUBO solver's goal between common work and this work.} 
\label{f:qubo_overview}
\end{figure}

To tackle this computational challenge and achieve near-optimal beamforming performance with a massive number of PIN-diode phase shifters, we study the use of physics-inspired computational technologies whose optimization principles are non-conventional and known for potential speedup~\cite{aramon2019physics,inagaki2016coherent,kim2019leveraging}. We consider physics-inspired optimization solvers for quadratic unconstrained binary optimization (QUBO) problems for which novel hardware is developed such as quantum annealers and Coherent Ising machines.
% In our early work, we  that quantum annealing can solve the
Unlike common QUBO solver approaches, where an NP-hard optimization problem is translated into a QUBO problem, our optimization problem of interest is expressed as a series of QUBOs, combined with a bisection method. Here, each QUBO in the series aims to check if the global optimum value is negative or not (\emph{cf.} finding the global optimum in common cases). This is a unique goal that motivates us to devise a new solver design with stopping criteria, which we believe could be a new way of using QUBO solvers:
Figure~\ref{f:qubo_overview} highlights this difference. 
In an earlier version of this work~\cite{stockley2023optimizing}, a \emph{quantum annealing} (QA) algorithm was used as a proof of concept, but this difference was not considered. 
% \textcolor{red}{Keith: I am not clear at all what is being claimed here. What is ``this difference''? And this paper should not give the impression that the method is new --- it is all in Stockley \& Briggs!}
Furthermore, we point out the limitations of the state-of-the-art quantum annealer as a solver for this distinctive QUBO structure later in the paper. 
% Instead, we use classical simulated annealing

The contributions of this work can be summarized as follows:

\begin{itemize}
    \item We base our work on the formulation of Oliveri et al.~\cite{Oliveri}. While the original formulation was limited to a half-space and vertical projection of a planar square patch on the sphere, we extend it fully into three dimensions and also choose $R$ to be a circular patch to consider more realistic beam propagation. 
    \item With respect to the optimization form, we suggest a procedure for converting the optimization objective function from a ratio of quadratic functions to a sequence of more easily solvable QUBO sub-problems. This conversion is an exact equivalence. 
    The optimization form for discrete 2- and 4-phase-shifting configuration is considered aiming to maximize the power radiated on a target area out of the total transmitted power.
    % Uniquely, the original optimization problem is expressed as a series of QUBO forms representing a bisection method. 
    % In the early version of this work, quantum annealing is used as a proof of concept~\cite{}, but we point out that current QA has limitations as a solver for the QUBO structure. 
    % The prior work considers a half-space and vertical projection of a planar square patch on the sphere. In our work, we extend it fully into three dimensions and use a circular patch for more realistic beam propagation.  
    \item We solve the QUBO structure using classical simulated annealing (SA), a generic QUBO solver. Considering the unique objective of each QUBO in the series in our formulation (Figure~\ref{f:qubo_overview}), we design stopping criteria for SA early stops. We observe that this early stop scheme is able to reduce compute time by about a factor of 10, without loss of beamforming performance. 
    % In our early work, we used quantum annealing, but we point out that current QA has limitations as a solver for the QUBO structure. 
    % \item
    % We believe this can suggest a new way of using Physics-inspired QUBO solvers for process acceleration. 
    \item We numerically evaluate the proposed scheme in comprehensive beamforming scenarios. We show that our formulation and solver design can efficiently increase beam gains as the sizes of the antenna array increase. 
    % It is observed that even discrete 4-phase shifting configurations can achieve similar performance to the optimal continuous setting. 
    We report up to a 10,000-element array with 4-phase cases showing its design scalability (note that only a 40-element array with 2-phase cases is borderline with a conventional integer programming solver). We observe the resulting discrete phase configuration via our optimization method with 4-phase shifting can achieve a similar beamforming performance to one that the optimal continuous setting can achieve.
    % , while the discrete one has relatively limited elevation angle ($\theta$) controllability.
    \item We show that this resulting discrete phase configuration cannot be estimated by quantizing continuous solutions. It is observed that quantized continuous solutions are efficient for relatively small array sizes, but they result in large undesired side-lobes as the array size increases. Thus, for large-scale antenna arrays with limited discrete phases, more advanced designs are required, and we demonstrate the feasibility of our proposed method.
\end{itemize}

% The recent work also studies QA for beamforming~\cite{ross2021engineering,lim2023quantum} for RIS.
Compared to prior discrete phase shifting beamforming work~\cite{wu2019beamforming,di2020hybrid,di2020practical,optimalbeamforming2022,zhang2022configuring}, this paper handles a more \emph{generic} transmitting-array 
synthesis problem in that the knowledge of receiver antenna structure and channel state information (CSI) is not assumed. This implies our design can be adapted to communication scenarios that work without CSI~\cite{souto2020beamforming,lai2023blind}.
Moreover, our scenarios are not limited to reconfigurable
intelligent surfaces (RISs)
and thus the optimization problem of interest is different. The recent work studies QA for beamforming with discrete phase-shifting configuration ~\cite{ross2021engineering,lim2023quantum}, but they are also discussed in the context of RISs. Finally, to the best knowledge, our work reports 3D beamforming performance with combinatorial optimization for extra-large arrays for the first time (up to 10,000-element scenarios). 
\

\section{Background}
\label{s:primer}

\subsection{QUBO Problems}
\label{s:ising_form}

The 
\emph{Quadratic Unconstrained Binary Optimization} (\emph{QUBO}, closely related to the Ising model of physics~\cite{ising1925beitrag}) 
is a discrete combinatorial optimization problem. Its state 
variables $x_i$ are bits (0 or  1), and the objective function (also called \emph{Hamiltonian}) is represented as a quadratic cost function of the following form with $x\in \{0,1\}^{N_V}$:
% $$
\begin{align}
\mathcal{H}(x) = \sum^{N_V}_{i\leqslant j} Q_{ij}x_i x_j,
\label{eqn:qubo}
\end{align}
% \begin{align}
% \hat{x}_1,\ldots,\hat{x}_N=
% \arg\min_{\left\{ x_1, \ldots, x_N \right\}} \sum^{N}_{i\leqslant j} Q_{ij}x_i x_j,
% \label{eqn:qubo}
% \end{align}
where $N_V$ is the number of bits, 
$\mathbf{Q} \in \mathbb{R}^{N_V
\times N_V}$ is upper triangular with elements being real values, and $N_V$ is the QUBO variable count that is $KN^2$.
The objective is to find the values of the binary variables that minimize this expression, that is, the  \emph{ground state}, which is the value of $\arg\min_{\left\{ x_1, \ldots, x_{N_V} \right\}} \mathcal{H}(x)$ and $\mathcal{H}(x)$ may be called the QUBO \emph{energy}, after the physics analogy. Solving QUBO problems can be challenging since as the problem size increases, finding the optimal solution becomes exponentially more difficult. However, there are various non-conventional physics-inspired methods and algorithms,
% , such as quantum annealing and classical optimization techniques like simulated annealing
which can be employed to solve QUBO problems and efficiently find near-optimal solutions. They are known for potential speedup over conventional methods due to their atypical optimization principles. With their promise, specialized hardware has recently emerged such as coherent Ising machines (CIM)~\cite{wang2013coherent,mcmahon2016fully}, oscillator Ising machines (OIM)~\cite{wang2019oim}, digital annealers~\cite{aramon2019physics}, and quantum annealers~\cite{kim2019leveraging}. Among many potential techniques, this work applies a generic simulated annealing algorithm that can be implemented on any platform as an initial guiding framework.

% In this work, we focus on the simulated annealing algorithm.
% However, note that we suggest a different way to use QUBO problems to deal 
% The off\hyp{}diagonal matrix
% elements $Q_{ij}$ ($i\neq j$) correspond
% to $g_{ij}$ in Eq.~\ref{eqn:ising}, and the diagonal
% elements correspond to $f_i$.

% \subsection{Bisection Method}
% \label{s:bisection}
% A quasiconvex optimization problem can be solved by a sequence of convex optimization problems. Assuming that the objective function $O(x)$ is quasiconvex, while $\mathcal{O}_t: \mathbf{R}^n \rightarrow \mathbf{R}, t \in \mathbf{R} $ be a family of convex functions that satisfy $f_o(x)\leq t \leftrightarrow \mathcal{O}_t(x) \leq 0$. If for each $x$, $\mathcal{O}_t(x)$ is a nonincreasing function of $t$, then $\mathcal{O}_s(x) \leq \mathcal{O}_t(x)$ whenever $s\leq t$.

% the bisection method.

\begin{figure}
    \centering
    \includegraphics[width=0.9\linewidth]{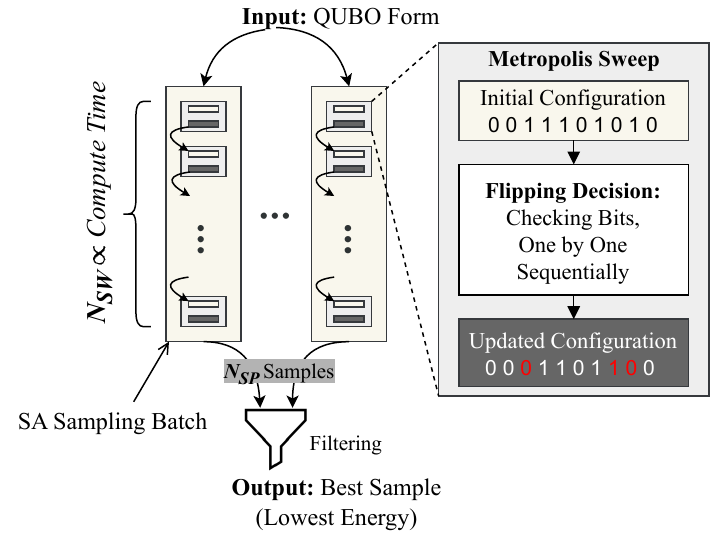}
    \caption{High-level SA system architecture, presenting typical optimization processing in SA.} 
\label{f:annealing}
\end{figure}

\subsection{Simulated Annealing}
\label{s:sa}

% , especially when dealing with large and complex search spaces. 
% The success of the algorithm depends on the choice of parameters, including the cooling schedule and the acceptance probability function. Properly tuning these parameters can lead to good results in practice, allowing Simulated Annealing to escape local optima and find near-optimal solutions to QUBO problems.

% \emph{PIC algorithms} are distinguished by their underlying principles used to find the global optimum of Ising problems (\emph{i.e.,} optimization principle). 
% A common part is they are \emph{probabilistic} 
% (Monte Carlo) heuristics and thus multiple independent runs of heuristics can improve the overall optimization performance. Further, each run has also a controllable factor that affects heuristics quality (\emph{e.g.,} iteration or time duration for searching)~\cite{kim2021physics,kim2019leveraging,aramon2019physics,inagaki2016coherent}. 

Annealing is a physical process of a material that changes its properties through heating and cooling. Being slowly cooled from high temperature, the material's atomic configuration converges to a low, possibly minimal, energy configuration~\cite{bertsimas1993simulated}. Inspired by the annealing process,
\emph{Simulated Annealing} (SA) is a rudimentary physics-inspired optimization algorithm that can be effective for finding near-optimal solutions to QUBO problems~\cite{aramon2019physics}.
Figure~\ref{f:annealing} presents the overall processing structure of a SA system. When a QUBO form of the problem of interest is formulated, it goes into each \emph{SA batch}. The batch's process (\emph{i.e.,} each sampling run) is designed to find the global optimum in ideal cases, which is based on Markov chain processes that mimic the convergence phenomenon of annealing as follows. 

With a QUBO problem, each batch prepares a random initial configuration (\emph{i.e.,} a random state) and starts \emph{Metropolis sweep} processing~\cite{metropolis1953equation}. 
In the sweep process, each bit of an initial configuration is sequentially explored (one by one), deciding whether its initial value is flipped to the other value (\emph{i.e.,} $0 \leftrightarrow 1$) or not. The flipping decision is probabilistic and its probability depends on the impact of QUBO energy difference, generally moving the given configuration towards the one with the lower energy. Assuming the $i$-th bit is being explored to decide the flip, the energy difference is calculated as:

\begin{equation}
\Delta{\mathcal{H}} = \mathcal{H}(x_0,x_1,\cdots,\bar{x_i},\cdots,x_{N_V}) - \mathcal{H}(x), 
% \text{with} \bar{x_i} = \begin{cases}
    % 1\,\,\,\, \text{if}\,\, x_i=0\\
               % 0\,\,\,\, \text{if}\,\, x_i=1
% \end{cases}
\end{equation} 

\noindent where $\bar{x_i}$ is the complement of $x_i$.  
If $\Delta{\mathcal{H}} <0$, the flip is always accepted. Otherwise, the system accepts it with the probability $p=e^{-\beta\Delta\mathcal{H}}$, with $\beta$ being the \emph{inverse temperature} $1/T$. Here the input temperature is an algorithmic parameter, and this probabilistic acceptance allows the algorithm to escape local optima. If the flip is accepted, an updated configuration becomes the next sweep's initial configuration and the sweep process iterates for $N_{SW}$ times (\emph{the number of sweeps}). The iteration can numerically simulate the annealing convergence process, finally thermalizing to the corresponding
temperature~\cite{metropolis1949monte}. The best configuration observed during the batch becomes the batch output. SA likely finds the ground state when with sufficiently large $N_{SW}$ in principle, but this could lead to extremely high compute time, which is not allowed in many applications. For this reason, $N_{SW}$ should be limited in reasonable ranges and then the SA algorithm becomes 
a probabilistic heuristic, acting like sampling a QUBO solution candidate. In this regard, the whole serial sweep process is often called \emph{sampling}, with each batch result being a sample. Naturally, $N_{SW}$ is one of the main factors that decide compute time and sampling quality. Typically, multiple batches are applied to increase the fidelity
of finding the global optimum, which can be run independently in parallel. In this work, 50 batches with $N_{SW}=50$ are used, by default. The SA batches are implemented in a sequential way, but we instead adopt stopping criteria (Sec~\ref{s:early_stop_design}). 

\section{METHODOLOGY}
\label{s:design}

We consider an abstraction of an antenna array as a collection (not necessarily planar) of isotropically radiating point elements.
These could represent active elements fed from cables, or passive elements used in either reflective or transmissive mode.
Our inspiration is the recent use of PIN diode arrays for reflecting or focusing millimeter-wave beams. 
These are currently opening up a cheap, simple, and reliable new technology with important new applications in propagating millimeter-wave signals beyond the usual line-of-sight restrictions.

\subsection{3D Beamforming Formulation.}
\label{s:ising_form_2p}

For antenna beamforming, the aim is to choose weights across antenna elements to maximize energy in a certain direction, allowing more accurate beamforming. The optimization problem of interest can be expressed as:

\begin{equation}
\hat{w} = \argmax{w}\frac{w^H A\; w}{w^H B\; w},
\label{eqn:objective}
\end{equation}
where $w$ is a complex vector (2-phase case: $w \in\{1,-1\}^{N^2}$ and 4-phase case: $w \in\{1,j,-1,-j\}^{N^2}$ for $N^2$ antenna elements), and $A$ and $B$ are Hermitian
positive-semidefinite matrices defined by surface integrals over $R$ (target surface) and $S$ (the whole sphere surface)
respectively. $A$ and $B$ are functions of the array geometry, beam azimuth and elevation, and beam angular diameter. Precisely,

\begin{equation}
A = \int_{R} T(x,y,z)T(x,y,z)^H \,dS \
\end{equation}
where $T(x,y,z)_i = \text{exp}\{(-jk(xX_i+yY_i+zZ_i)\}$ and $X_i, Y_i, Z_i$ are antenna element locations. $A$ and $B$ are always Hermitian positive-semidefinite matrices (\emph{i.e.}, $A=A^H$ and $z^HAz > 0 \,(\forall z \neq 0)$, and the same for $B$). 

If $w$ is unconstrained (\emph{i.e.,} \emph{continuous} phase and amplitude shifting), this maximization problem (Eq.~\ref{eqn:objective}) can be easily solved by finding the largest eigenvalue
$\lambda_\text{max}$ of the generalized eigenvalue problem $Aw=\lambda Bw$, and the corresponding eigenvector $w_\text{max}$ is the desired steering vector. 
% If $w$ is unconstrained (\emph{i.e.,} \emph{continuous} phase shifting), this can be solved by finding the largest eigenvalue $\lambda_{\text{max}}$ of the generalized eigenvalue problem, $Aw=\lambda Bw$. 
% The corresponding eigenvector $w_{\text{max}}$ is the optimal solution. 
The continuous setting is able to apply the optimal solution, but it requires complicated hardware. On the other hand, when each element of $w$ needs to be decided as one of only a few possible discrete phases, \emph{e.g.,} $\{1,j,-1,-j\}$ (equivalently 0, 90, 180, 270 degrees), the hardware implementation becomes simple and straightforward, but finding the best solution among the possible configurations (\emph{i.e.,} constrained \emph{discrete} phase shifting) becomes an NP-hard problem. We use physics-inspired optimization methods in order to solve this hard problem more efficiently. Note that the objective in the equation is the ratio of two positive convex functions, which implies a bisection method is applicable~\cite{boyd2004convex}. The bisection method for Eq.~\ref{eqn:objective} can be expressed as a series of QUBO forms that can be solved by physics-inspired optimization methods (as discussed above in Sec~\ref{s:primer}). 
% As proof of concept, we use a quantum annealing (QA) algorithm in prior work. However, we point out the current QA's limitations for the approach (Sec~\ref{s:need_of_early_stop}), and instead, we delve into a generic SA algorithm (Sec~\ref{s:sa} with stopping criteria based on the goal and principle of the series of QUBO forms for the bisection method. 
% For the rest of the subsection, 
We now introduce the bisection method and then the corresponding QUBO formulation to Eq.~\ref{eqn:objective} in the context of the bisection method.

% In this subsection, we introduce 

\parahead{Bisection method.} Let us first consider a general form, with positive convex functions $f(x), g(x)$, both mapping $\R^n$ to $\R$, and $C$ some constraint set (typically $C=\{0,1\}^n$):
\begin{equation}
\max_{x\in C}\frac{f(x)}{g(x)}.\label{eqn:feas}
\end{equation}
This is precisely equivalent to
\begin{equation}
\min_{t\in\R} t\quad \text{such that}\quad f(x)-tg(x)< 0\quad \forall\ x \in C.
\label{eqn:equiv}
\end{equation}
To start the bisection, we first need to bound the optimal $t$.  For the antenna problem, the initial bounds $(t_0,t_1)=(0,1)$ will always work; we will be defining $f(x)=x^T Ax$ and $g(x)=x^T Bx$ with $A$ defined by an integral over a smaller region than $B$, and so the optimal $t$ cannot be greater than unity. One bisection step then involves setting a trial $t$ half-way between current the bounds $(t_0,t_1)$, and solving the minimization problem
\begin{equation}
\min_{x \in C}  f(x) - tg(x).
\end{equation}
If the optimal value of the expression is positive, then it follows that the trial $t$ is too small, so we can raise the lower bound to $t_0=t$. Conversely, if the optimal value is negative, we set $t_1=t$. Bisection can be stopped when the interval $(t_0,t_1)$ is sufficiently small ($10^{-6}$ in our implementation). Note that during the minimization process, the finding of \textit{any} positive value for $f(x)-tg(x)$ means that the constraint in (\ref{eqn:equiv}) is violated, and this allows us to abort the minimization early and make the correct decision $t_0=t$ (see Sec~\ref{s:early_stop_design} for algorithm details).  
% This means that a heuristic minimization method can be safely used for updating the lower bound.
This is, however, not the case for upper-bound updates, so these steps, whether done by integer linear programming or otherwise, are typically slower. To accelerate this process, we also suggest another stopping criterion for the upper-bound updates (Sec~\ref{s:early_stop_design}).

\parahead{QUBO formulation.}
In order to make use of physics-inspired QUBO solvers (such as SA, QA, and other Ising machines), the optimization variable $x$ must be a bit-vector, containing only the values 0 or 1. We map these to complex weights by an affine map; in the 2-phase case this takes the form $w_i=\alpha x_i+\beta$ for appropriate complex constants $\alpha$ and $\beta$.  In the \mbox{4-phase} case the map is similar, but uses two bits for each element of the weight vector. The optimization problem now has the following slightly more general form for an antenna array with $n$ elements, and the bisection method still applies. In this form all parameters ($a,c\in\R^n$, $b,d\in\R$), variables, and function values are now real:
\begin{equation}
\max_{x\in \{0,1\}^n} \frac{x^T A\; x +a^T x+b}{x^H B\; x +c^T x+d}.
\end{equation}
For this objective, the specific form of Eq.~\ref{eqn:feas}, which forms the constraint in Eq.~\ref{eqn:equiv}, is now
\begin{equation}
\min_{x\in \{0,1\}^n} x^T (A-tB)\; x + (a-tc)^T\; x  +b-d,
\end{equation}
\noindent which is a QUBO consisting of quadratic and linear terms. Note that a linear terms can be absorbed into the diagonal elements of $A$, since $x_i^2=x_i$ for 0-1 vectors.  
The constant terms defined by $b$ and $d$ are simple shifts which make no essential difference to the solution procedure.

% \textcolor{red}{above vs. below one for explanation (or connection?)}
%Our end goal is to evaluate the function:

% \begin{align}
% % \centering
% B=-2\pi jA/\lambda \\
% g(u,w)=w\cdot \text{exp}(Bu) \\
% G(u,w) = |g(u,w)| \\
% q_0 = \max_{x\in C} {f(x)-t_0g(x)}] \
% = -\min_{x\in C} {t_0g(x)-f(x)}.
% \end{align}

One step of the bisection procedure is to decide whether the following inequality is satisfiable:

\begin{equation}
(x^TA_0x + {b_0}^Tx+c_0) -t(x^TA_1x + {b_1}^Tx+c_1) < 0.
\end{equation}
This can be decided by solving the subproblem:

\begin{equation}
q_1 = \max_{x\in C} {f(x)-t_1g(x)} \
= -\min_{x\in C} {t_1g(x)-f(x)}.
\end{equation}
If $q_1<0$, $t_1$ is too large, and $t_*<t_1$. Once we have checked that our initial bounds $(t_0, t_1)$ are such that $t_0<t_*<t_1$, we are ready to start our bisection steps in order to find $t_*$.

We begin with $t=\alpha t_0 + (1-\alpha)t_1$, where $\alpha \in (0,1)$. For $q$ as follows:

\begin{equation}
q = \max_{x\in C} {f(x)-tg(x)} 
= -\min_{x\in C} {tg(x)-f(x)}.
\end{equation}
% if $q>0$, then we know that $t$ is too small, so we set $t_0=t$. If, however, $q<0$, then we know that 
\begin{equation}
Q = \min_{x\in C} {tg(x)-f(x)}.
\end{equation}

% \begin{equation}
% \max_w \frac{w^\dagger A\; w}{w^\dagger B\; w}.
% \end{equation}

For Eq.~\ref{eqn:objective}, $\mathbf{w} = (w_1-w_0)z + w_0 \mathbf{1}$ with $\mathbf{1}$ being the $n$-dim vector of all 1's and $z \in(0,1)^n$ (\emph{i.e.,} $w=w_0$ with $z=0$, while $w=w_1$ with $z=1$). Let $\Delta=w_1-w_0$, with $w_0, w_1$ being the two phases we want to transmit at. We can then evaluate $w^\dagger Aw$ using this substitution:

\begin{small}
\begin{align}
w^\dagger Aw &= (\Delta^{*}z^T + {w^*_0}\mathbf{1}^T)A(\Delta z + w_0\mathbf{1}) \notag\\
% = \Delta^{*}z^TA\Delta z + A\Delta^{*}z^{T}A(w_0\mathbf{1})+(w^*_0\mathbf{1}^T)A\Delta z +(w^*_0\mathbf{1}^T)A(w_0\mathbf{1})
% \notag\\
&=|\Delta|^2z^TAz+2\mathbb{R}(\Delta^*w_0z^TA\mathbf{1})+|w_0|^2\Sigma_{i,j}A_{ij} \notag\\
&=|\Delta|^2z^TAz+2\mathbb{R}(\Delta^*w_0\mathbf{1}^TAz)+|w_0|^2\Sigma_{i,j}A_{ij}
\end{align}
\end{small}

We can follow a similar process for $w^\dagger Bw$, and thus we retrieve the final forms for our coefficient forms from the given data alone:

\begin{small}
\begin{align}
A_0 &= |\Delta|^2A \\
b_0 &= 2\mathbb{R}(\Delta^*w_0\mathbf{1}^TA)\\
c_0 &= |w_0|\Sigma_{i,j}A_{ij},
\end{align}
\end{small}

and similarly,

\begin{small}
\begin{align}
A_1 &= |\Delta|^2B \\
b_1 &= 2\mathbb{R}(\Delta^*w_0\mathbf{1}^TB)\\
c_1 &= |w_0|\Sigma_{i,j}B_{ij}.
\end{align}
\end{small}

% With $w=\Delta z + w_0\mathbf{1}$, $w=w_0$ with $z=0$, while $w=w_1$ with $z=1$.

Until now, we have handled the 2-phase case of our antenna beamforming problem (\emph{i.e.}, the case where each antenna element can transmit with one of two phases). While this maps more naturally to our formulation of the optimization vector as a Boolean vector, allowing only 2-phase transmission massively decreases not only with the maximum gains of the beams we are producing, but the accuracy with which we can steer our beam as well.

\parahead{4-Phase extension.}
% \label{s:ising_form_4p}
Therefore, we want to extend the method we used to determine our linear coefficients previously to the 4-phase case.
Following similar steps using 
% $w=au+bv+c\mathbf{1}$ and 
$w^\dagger Aw= (a^*u^T+b^*v^T+c*\mathbf{1}^T)A(au+bv+c\mathbf{1})$ with $u,v$ being a Boolean vector, the 4-phase case results in
% \noindent{}where 

% \begin{align*}
% w^\dagger Aw= (a^*u^T+b^*v^T+c*\mathbf{1}^T)A(au+bv+c\mathbf{1})\\
% =(a^*u^T)A(au) + (a^*u^T)A(bv)+(b^*v^T)A(au)+(b^*v^T)A(au)\\+(b^*v^T)A(bv)
% +(a^*u^T)A(c\mathbf{1})\\+(b^*v^T)A(c\mathbf{1})+(c^*\mathbf{1}^T)A(au)+(c^*\mathbf{1}^T)A(bv)\\
% +(c^*\mathbf{1}^T)A(c\mathbf{1})\\
% =|a|^2u^TAu+|b|^2vTAv+2\mathcal{R}\left(a^*u^TA(bv)\right)\\
% +a^*cu^TA\mathbf{1}+ac^*\mathbf{1}^TAu+b^*cv^TA\mathbf{1}+c^*b\mathbf{1}^TAv\\
% +|c|^2\mathbf{1}^TA\mathbf{1}
% = |a|^2u^TAu + |b|^2v^TAv+2\mathcal{R}\left(a^*u^T)A(bv)\right)\\
% +2\mathcal{R}\left(a^*c(u^TA\mathbf{1})\right)+2\mathcal{R}\left(b^*c(v^TA\mathcal(1))+|c|^2\mathbf{1}^TA\mathbf{1}\right)\\
% = |a|^2u^TAu+|b|^2v^TAv +u^T\mathcal{R}\left(a^*bA\right)v\\
% +2u^T\mathcal{R}\left(a^*cA\right)\mathbf{1}+2v^T\mathcal{R}\left(b^*cA\right)\mathbf{1}+|c|^2\mathbf{1}^TA\mathbf{1}\\
% =\begin{bmatrix} 
%     u \\
%     v
% \end{bmatrix}^T \begin{bmatrix} A_{00} & A_{01} \\
% A_{10} & A_{11}
% \end{bmatrix}
% \begin{bmatrix} 
%     u \\
%     v
% \end{bmatrix}
% + \text{linear terms} + \text{constant terms},
% \end{align*}

\begin{small}
\begin{align}
A_0 &= \begin{bmatrix} A_{00} & A_{01} \\
A_{10} & A_{11}
\end{bmatrix}\\
b_0 &= 2\mathbb{R}(a*c\mathbf{1}^TA)+2\mathbb{R}(b*c\mathbf{1}^TA)\\
c_0 &= |c|^2\Sigma_{i,j}A_{ij}, 
\end{align}
\end{small}

\noindent and similar for $A_1, b_1, c_1 \text{ with } B$, where $A_{00}=|a|^2A$, $A_{01}=a^*bA$, $A_{10}=(a^*bA)^\dagger$ and $A_{11}=|b|^2A$.
Then the four phases are $w_0=c$, 
% (with $u=0,v=0$), 
$w_1=b+c$, 
% (with $u=0,v=1$), 
$w_2=a+c$, 
% (with $u=1,v=0$), 
and $w_3=a+b+c$. 
% (with $u=1,v=1$). 
With $[a,b,c] = [-j-1,j-1,1]$, then $w=[1,j-j,1]$.

\parahead{Energy difference computation.} For the sweep process in SA, the energy difference between the current configuration ($x$) and a bit-flipped configuration ($\bar{x}$) needs to be calculated per variable. Thus, the bulk of the computation time in SA is taken up in calculating the energy change after a particular flip occurs. For this reason, we utilize a \emph{difference function} that pre-computes the coefficients for the quadratic form, to save some computation time. 

% The energy of the QUBO objective function is:

% \begin{equation}
% \mathcal{H}(\mathbf{x}) = \mathbf{x}^T\mathbf{A}\mathbf{x}=\Sigma_{i,j} A_{ij}x_ix_j
% \end{equation} 

% \noindent{}with $\mathbf{x}\in \{0,1\}^n$. 

Generally in QUBO, the energy difference between $x$ and $\bar{x}$ can be expressed as:
% \begin{equation}
% \Delta{\mathcal{H}} = \mathcal{H}(x_0,x_1,\cdots,\bar{x_i},\cdots,x_n) - \mathcal{H}(\mathbf{x})
% \end{equation} 
\begin{small}
\begin{align}
\Delta(x,k)&= (\bar{x}^TA\bar{x}+b^T\bar{x}+c) - (x^TAx+b^Tx+c)\notag\\
&=(\bar{x}^TA\bar{x}-x^TAx)+(b^T\bar{x}-b^Tx)
\end{align}
\end{small}

\noindent{}where $\bar{x}$ is $x$ with $x_k$ ($k$-th component) flipped. Accordingly, component-wise:
\begin{equation}
\left(\Sigma_{i,j}\bar{x_i}A_{ij}\bar{x_j}-\Sigma_{i,j}x_iA_{ij}x_j\right) + \left(\Sigma_i b_i\bar{x_i}-\Sigma_i b_ix_i\right).
\end{equation}
Therefore, quadratic terms are
\begin{multline}
\Sigma_{i,j} \bar{x_i}A_{ij}\bar{x_j} - \Sigma x_iA_{ij}x_j\notag\\
=\Sigma_{i=k\neq k} (\bar{x}_k-x_k)A_{kj}x_j+\Sigma_{j=k\neq i} x_iA_{ik}(\bar{x}_k-x_k)\notag\\+(\bar{x}_k-x_k)A_{kk},\notag
\end{multline}
and the linear terms are
\begin{equation}
    \Sigma_i b_i\bar{x}_i = \Sigma_{i\neq k} b_i\bar{x}_i - b_ix_i + b_k\bar{x}_k - b_kx_k = b_k(\bar{x}_k-x_k).\notag
\end{equation}
We define
\begin{align}
    \delta x_k := \bar{x}_k-x_k = \begin{cases}
    \hphantom{-}1\,\,\,\, \text{if}\,\, x_k=0\\
               -1\,\,\,\, \text{if}\,\, x_k=1
    \end{cases},\notag
\end{align}
and therefore we get the difference function:
\begin{multline}
    % \begin{split}
    \Delta(x,k) = \Sigma_{i=k \neq j} \delta x_k A_{kj}x_j \notag\\ 
    + \Sigma_{j=k \neq i} \left( x_iA_{ik}\delta x_k + \delta x_kA_{kk}+b_k\delta x_k\right).
    \label{}
    % \end{split}
    % \label{eqn:difference_function}
\end{multline}

% \begin{align}
% \begin{small}
% \mathbf{M} = \begin{bmatrix} 
%     f_{1} & g_{12} & \dots & g_{1N_V} \\  0 & f_{2} & \ddots & \vdots  \\
%       \vdots     & \ddots  &  \ddots & g_{(N_V-1)N_V} \\
%     0       & \dots & 0 & f_{N_V} 
%     \end{bmatrix}.
% \notag
% \label{eqn:ising_matrix}
% \end{small}
% \end{align}

\subsection{SA Early Stop and Stopping Criteira}
\label{s:early_stop_design}

% NEED TO COMPLETE THE SUBSECTION

The research question we now address is how to accelerate SA processing without loss of performance. 
If possible, this would be a great benefit, since systems can retrieve compute resources from SA earlier, and those extra available resources can be utilized for other processes.

% Unlike conventional Ising formulations where optimization forms are equivalently expressed as Ising forms, our Ising formulation

\parahead{Need of early stop.}
The goal of SA for the optimal performance per QUBO instance is to collect the ground state (global optimum) at least once among $N_{SP}$ collected samples. In each sampling process, not just an SA batch, but each sweep processing as well is a probabilistic heuristic optimization component. This implies any points of sweep iteration can find the ground state. In lucky cases, the first or second sweep (early at iteration) could happen to find the ground state, even if the systems' $N_{SW}$ setting is over a few tens. Once the ground state with the minimum possible QUBO energy is found, the rest of the sweeps will likely not lead to any bit flips, wasting compute time and resources, since flipping is generally applied towards the lower energy (even though probabilistic flips still occur). Therefore, leveraging the sequential structure and probabilistic properties of SA heuristics, its optimization processing can be further accelerated by an early stop, whenever the ground state is found in the middle
of sweep iterations (\emph{i.e.,} early convergence).

In this regard, automated early stop schemes are potential strategies that can accelerate the SA process. Here, for each QUBO instance, a SA system is expected to terminate the processes early, when the global optimum is found in the middle of sweep iterations. A clear challenge here is that systems in practice do not know the ground state of problems, so they do not know when to stop and terminate all the processes. Thus, designing a stopping criterion that guides when SA processing should stop, implicitly indicating the unknown ground state is found while it needs to be differentiated from being stuck in local minima, is a challenge. Interestingly, our QUBO formulation has an additional straightforward stopping criterion due to its unique objective as mentioned in the previous subsection. Furthermore, each QUBO in the series is much simpler than the original optimization problem, where the early stop scheme could effectively accelerate the overall optimization process.

% toward real-time processing

% Particularly, considering up to only a few milliseconds of extremely tight and strict latency deadlines for real-time processing in baseband systems and wireless networks (which are not commonly observed in other areas), accelerating PIC is an attractive research direction in PIC for PHY.

% clear criterion (further computation is meaningless)

\begin{figure} % ~/beam_forming-2.0> bash plots_for_EuCAP_paper.sh
\centering
% \subfigure
{\includegraphics[width=0.83\linewidth]{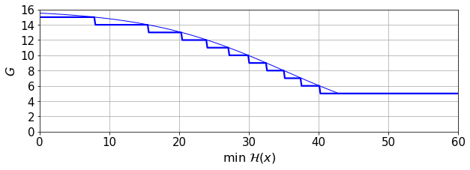}}\vspace{-0.15cm}
% \subfigure
\caption{Maximum batch iteration ($G$) for a stuck configuration ($x$) depending on the recent best min $\mathcal{H}(x)$.\vspace{-0.25cm}}
\label{f:max_batch}
\end{figure}

\begin{figure} % ~/beam_forming-2.0> bash plots_for_EuCAP_paper.sh
\centering
% \subfigure
{\includegraphics[width=0.22\textwidth]{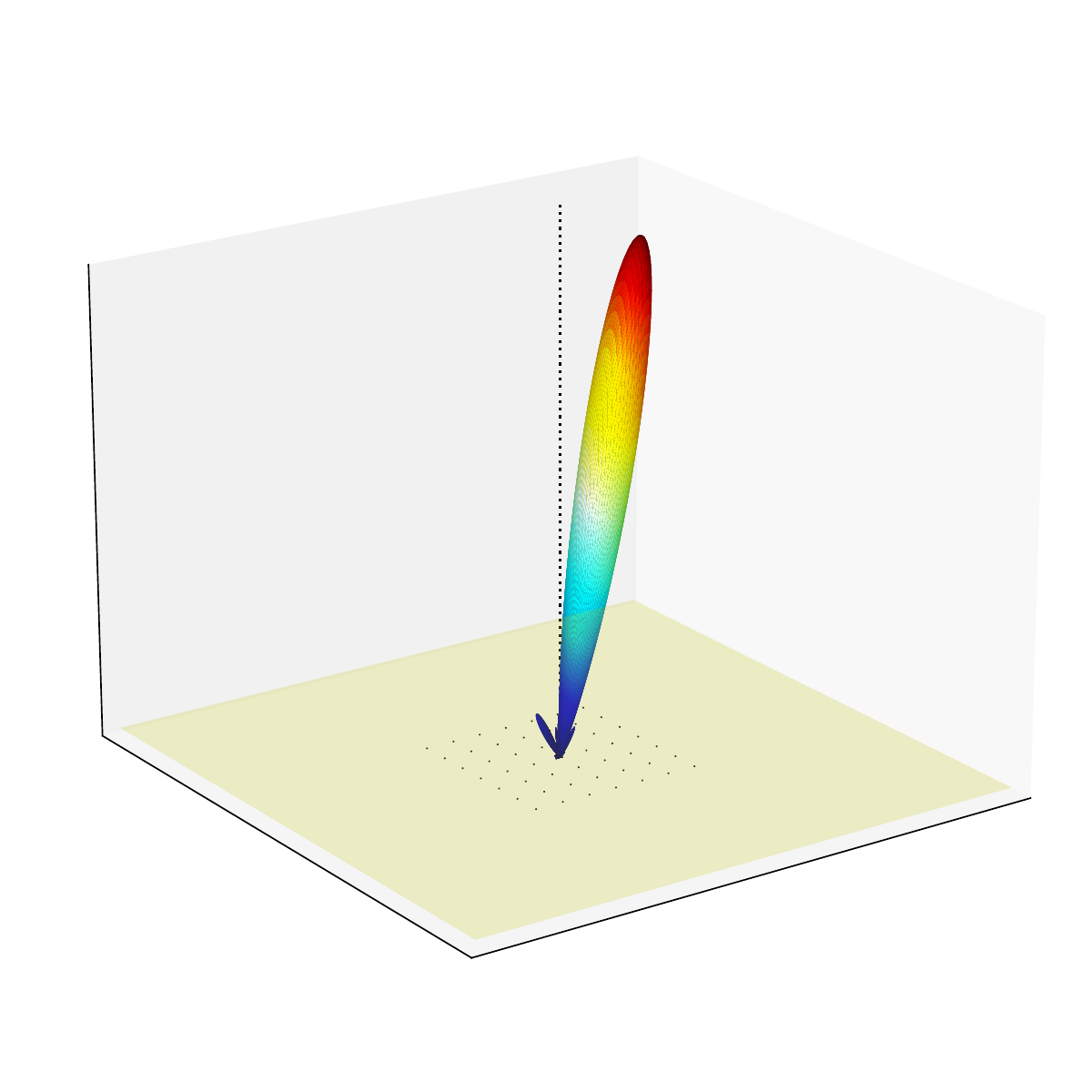}}\vspace{-0.15cm}
% \subfigure
{\includegraphics[width=0.22\textwidth]{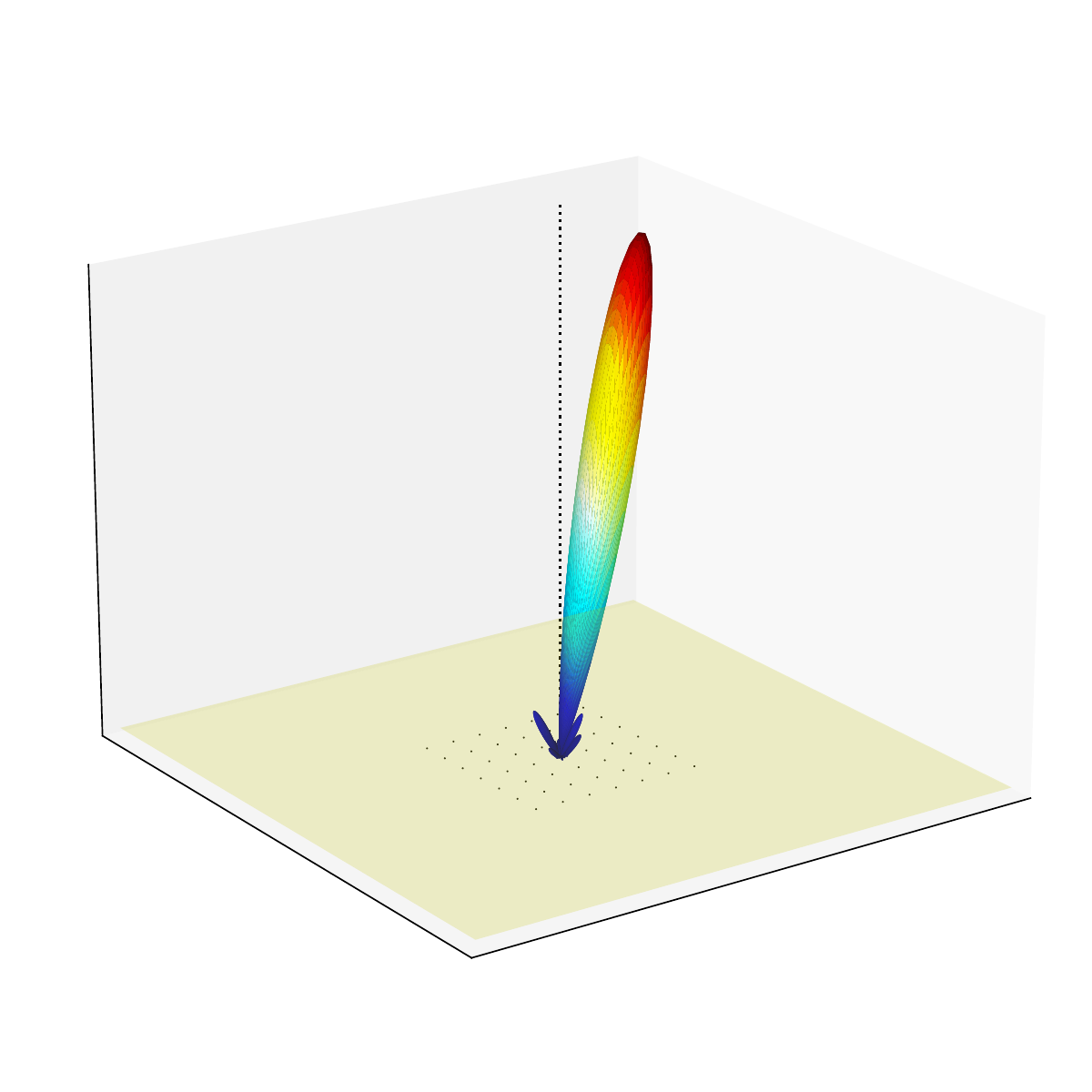}}
\caption{Proof of concept (the figure directly from the early version of this work~\cite{stockley2023optimizing}). Beamforming using a 7$\times$7 array of 4-phase elements, with backplane, azimuth=2 ($\phi$ = 114.64 degrees), polar elevation angle=0.25 ($\theta$ = 14.3 degrees). The phases are $\{1,j,-1,-j\}$ or equivalently 0, 90, 180, 270 degrees. The problem is solved by SA (\emph{left}), while the same problem is solved by QA on the D-Wave quantum annealer (\emph{right}). 
% \textcolor{red}{If you didn't actually run D-Wave yourself, then you have to say that this figure was taken from Stockley-Briggs.}
% \textcolor{red}{NB: state that these results are taken directly from Stockley \& Briggs!}
The small differences are due to the heuristic nature of the annealing process, meaning that slightly different (but still close to optimal) weights are found.}
\label{f:dwave_7x7}
\end{figure}

% \label{s:early_stop_design}
% \parahead{Stopping Criteria.}
% desired steering vector. 

\begin{small}
\begin{algorithm}  
\caption{SA Process w/ Stopping Criteria for Early Stop}   
\label{al:early_stop_algorithm}
\begin{algorithmic}
% \State Split \textbf{X} (D-dimension) data set into K sets.
\For{each batch}
        \State Prepare a random initial QUBO configuration ($x$)
        \For{each Metropolis sweep}
        \For{each QUBO variable $x_i$}
        \State Flip the bit ($\bar{x}_i$)
        \State // Check early stop feasibility
        \If{$\mathcal{H}(x_0,x_1,\cdots,\bar{x_i},\cdots,x_n) < 0$}
        \State
        Stop the process (\textbf{C1})
        \Else
        % \State // Check \textbf{C2} 
        \If{No update in min $\mathcal{H}(x)$ for $G$ batches}
        \State
        Stop the process (\textbf{C2}) // 
        % $G$ is decided by the function 
        (\emph{e.g.,} high $G$ when min $\mathcal{H}(x)$ is close to zero).
        \EndIf
        \EndIf
        \State // Flipping decision
        \State // $\Delta{\mathcal{H}} = \mathcal{H}(x_0,x_1,\cdots,\bar{x_i},\cdots,x_n) - \mathcal{H}(x)$
     
        \State Calculate energy difference ($\Delta \mathcal{H}$)
        % \State
           \If{$\Delta\mathcal{H}<0$}
        \State Accept the flip
        \Else
        \State Accept the flip with probability $p= e^{-\beta\Delta\mathcal{H}}$
        \EndIf
        \EndFor
        \State Record the update ($x$, min $\mathcal{H}(x$))
        \EndFor
        % \State PCA.Fit( // Fit PCA to Train Set.
        % \State
        % \State // Apply PCA transform to both Train Set and Test Set.
        % \State PCA.FitTransfrom
        % \State PCA.FitTransfro
        % \State 
        % \State // Apply inverse PCA transform to both Train Set and Test Set.
        % \State nsform(
        % \State
        % InverseTransform
        % \State
        % \State Compute MSE between 
%         \EndWhile
\EndFor
% \State Choose the best min $\mathcal{H}$ 
\end{algorithmic}    

\end{algorithm}
\end{small}

\begin{figure*}[ht] % ~/beam_forming-2.0> bash plots_for_EuCAP_paper.sh
% \centering
% \begin{minipage}[b]{1.0\linewidth}
% \centerline{\includegraphics[width=0.8\linewidth]{Figure/beam_pattern_continious.png}}
% \centerline{(a) Continuous Phase}
% \label{f:spectral}
% \end{minipage}

\begin{minipage}[b]{0.7\linewidth}
\includegraphics[width=0.95\linewidth]{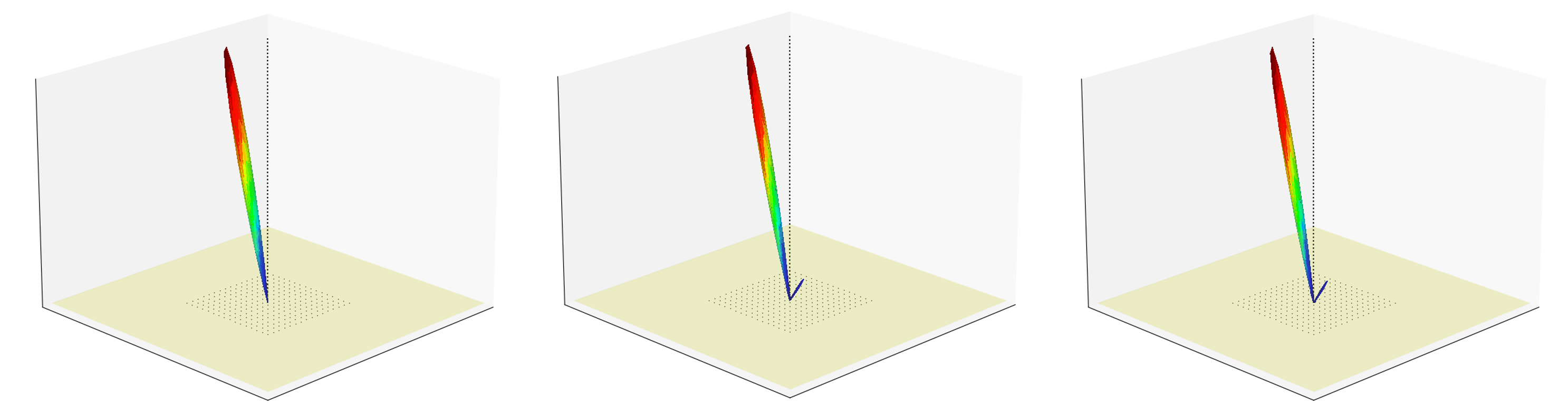}
\centerline{(a) $\mathbf{16\times 16}$ Array ($16^2$ Elements).}
\label{f:spectral}
% \end{minipage}

% \begin{minipage}[t]{0.7\linewidth}
% \centering
\includegraphics[width=0.95\linewidth]{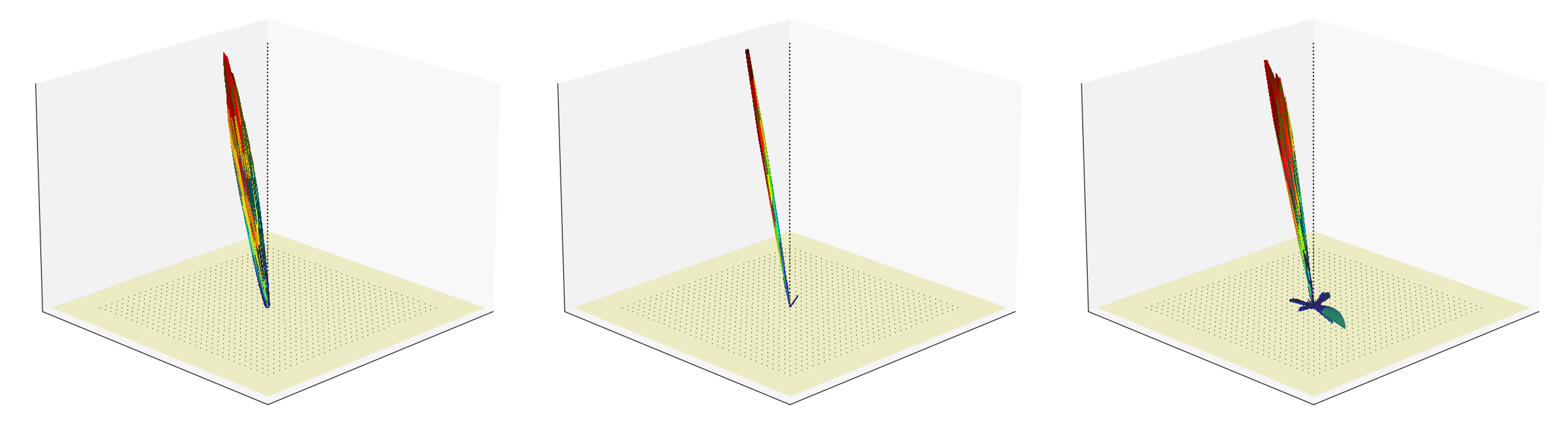}
\centerline{(b) $\mathbf{32\times 32}$ Array ($32^2$ Elements).}
\label{f:spectral_2}
\end{minipage}
\hfill
\begin{minipage}[b]{0.3\linewidth}
\centering
\includegraphics[width=0.98\linewidth]{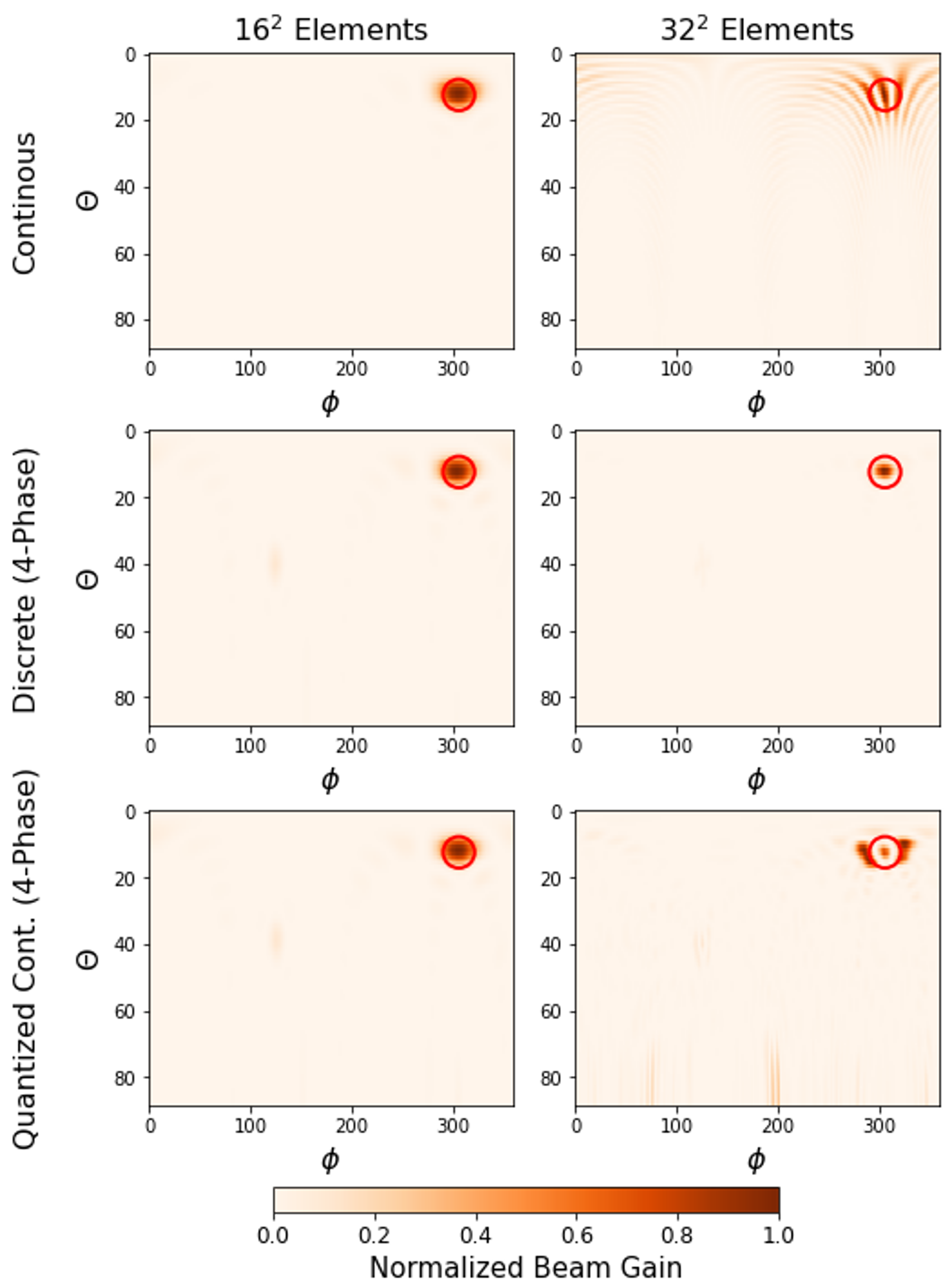}
\centerline{(c) Beam gain heatmap.}
\label{f:normalized_beam_heatmap}
\end{minipage}
\vfill

% \begin{minipage}[b]{1.0\linewidth}
% \centerline{\includegraphics[width=0.8\linewidth]{Figure/beam_pattern_quantized_cont_4p.png}}
% \centerline{(c) Quantized 4 Phases.}
% \label{f:spectral}
% \end{minipage}
\caption{\textbf{(a) \& (b):} Beamforming comparison (target $\theta = 12.38^{\circ}, \phi=306.16^{\circ}$) with different phase-shifting approaches: (\emph{left}) optimal continuous, (\emph{mid}) discrete (4-phase), and (\emph{right}) quantized continuous (4-phase) shifting. The max beam lengths are normalized by the max gains. 
\textbf{(c):} Normalized beam gain heatmap for the phase shifting methods. The (red) circles highlight the target beam direction. Two array scenarios are considered: $16\times16$ and $32\times32$ cases. 
% (actual gain comparisons are in Figure~\ref{f:gain_across_elements}).
}
\label{f:3D_benchmark}
\end{figure*}

\parahead{Proposed stopping criterion for early stop.} In this regard, for our design to reduce overall computation time for optimizing discrete phase configuration and to make the WPT beamforming system more efficient, we suggest two stopping criteria for an early stop in the SA processing based on the principle of the bisection method (Sec~\ref{s:ising_form_2p}).

The overall SA processing with our suggested stopping criterion is stated in Algorithm~\ref{al:early_stop_algorithm}.
% , the system has a quite straightforward stopping criterion for the early stop. 
Recall that, unlike common QUBO approaches, the goal of solving our QUBO form in the series for the bisection method is to find whether the optimal (minimum) value of the objective function is negative or not (\emph{cf.} finding the ground state in most applications). This implies whenever the sweep process finds a negative value, it can guarantee that the optimal value is also negative, and thus it is the point from which any further sweeps or iterations are not necessary. Therefore, $\mathcal{H}<0$ can be a safe stopping criterion (we call it \textbf{C1}).
% that does not affect the end result of the corresponding Ising form. 
Note that this is an uncommon and straightforward stopping criterion (uniquely observed in the bisection QUBO form due to its distinctive goal) that is \emph{always valid} regardless of whether the current state $x$ is the actual ground state or not.

The second criterion is designed to stop the sweep process early when it seems clear that the optimal value will likely not be negative. In some cases, large values of the QUBO objective function keep being obtained (\emph{e.g.,} over 100), despite further sweep iteration. Then it is likely that the optimum value is also non-negative and thus the system can stop the process. The challenge here is that unlike \textbf{C1} the probability of a wrong decision exists since the system does not know the exact global optimum. Especially when $\mathcal{H}(x)$ tends closer to 0 as further sweep and batch iterations are applied, we need a more conservative criterion. For this, we set a function that decides the maximum batch iteration ($G$) for a stuck configuration (\emph{i.e.,} no min $\mathcal{H}(x)$ update), by modifying a sigmoid function so that the current energy is considered for $G$ and the early stop (\textbf{C2}); in the function, we range $G$ between the minimum of five and the maximum of 15 as shown in Figure~\ref{f:max_batch}. Intuitively, for large min $\mathcal{H}(x)$, small $G$ is applied, while for small positive min $\mathcal{H}(x)$, high $G$ is applied. Whenever the system finds the better $\mathcal{H}(x)$, it resets $G$ based on the function to consider both the current energy and the energy-reducing trend. Note that even in a batch, the QUBO energy is checked for every variable, which implies at least $5N_V$ update trials (\emph{e.g.,} over 2,000 trials for 4-phase 256 elements) will be applied to make sure that a sufficient search process has been conducted for optimization.\vspace{-0.35cm}
% , to make sure that a meaningful optimization search process has been conducted.
% NEED TO EXPLAIN THE FUNCTION FURTHER

% In the early version of our work, 
\parahead{Limitation of current QA for early stop.}
In the early version of this work~\cite{stockley2023optimizing}, the feasibility of using physics-inspired optimization for the discrete phase-shifting configuration was demonstrated, as a proof of concept. 
% In previous work \cite{stockley2023optimizing}, the authors demonstrated the feasibility of using physics-inspired optimization for the discrete phase-shifting configuration. 
Figure~\ref{f:dwave_7x7} (which is taken directly from \cite{stockley2023optimizing}) shows that even a 4-phase discrete configuration can steer the beam toward a target direction precisely, solving the QUBO structure with both the SA (\emph{left}) and QA (\emph{right}) methods. Here, a very much simplified SA algorithm was used where the fixed probability $p$ was applied with fixed batch sizes and sweep counts. In other words, the solver design was not optimized given the unique objective of QUBO (Sec~\ref{s:ising_form_2p}). Furthermore, in the case of QA, optimization algorithms are based on adiabatic quantum computing principles which require high coherence, and thus end-users cannot intervene in the middle of the computation~\cite{kim2022warmstarted}. This implies the system cannot check an intermediate state required to decide whether the system can stop its search processing (\emph{i.e.,} an early stop scheme is not applicable in QA). Alternatively, sequential iterative QA runs could be considered, but very large overheads will be caused by multiple QA runs. For example in the early version, it is observed that the embedding process that maps a QUBO form into a sparse hardware graph takes most of the computing time. Therefore, current QA approaches are \emph{not} quite suitable to address our problem of interest, but advanced designs could be considered in the future aiming to address a series of bisection QUBOs and early stops.

\begin{figure}
\centering
\includegraphics[width=\linewidth]{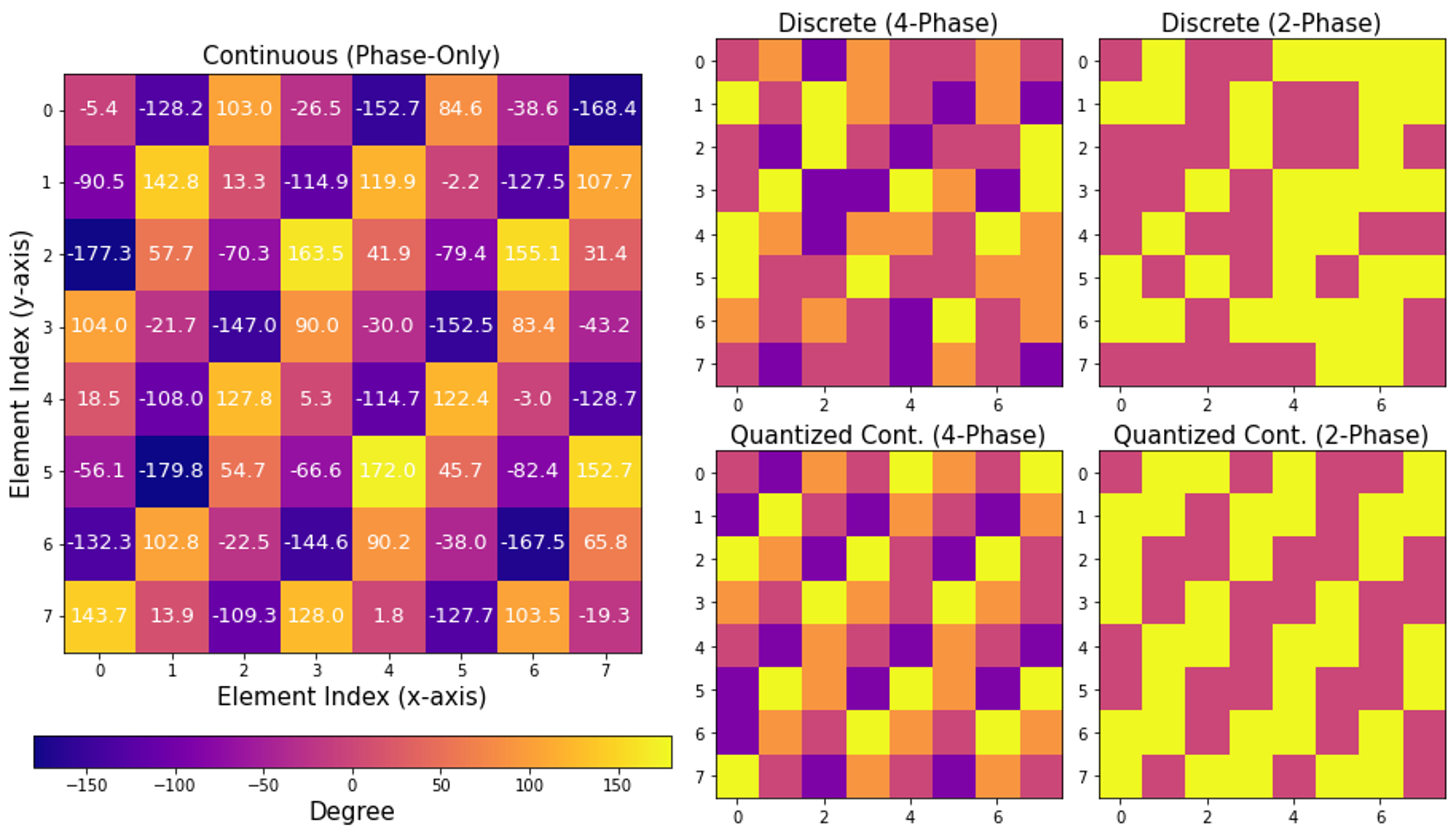}\vspace{-0.15cm}
\caption{$8\times8$ array configurations of continuous phase and discrete 
 2-phase (0, 180 degrees) and 4-phase (0, 90, 180, 270 (-90) degrees) shifting with our method (\emph{upper}) and quantized continuous one  (\emph{lower}).} 
\label{f:phase_configuration}
\end{figure}

% \begin{figure}
% \centering
% \includegraphics[width=0.7\linewidth]{Figure/normalized_beam_1.png}
% \caption{Normalized beam gain heatmap for different phase shifting methods: continuous weight, quantized 4-phase shifting, and discrete 4-phase shifting. The (red) circles highlight the target beam direction. Two array scenarios are considered: $16\times16$ (\emph{left}) and $32\times32$ (\emph{right}) cases.} 
% \label{f:normalized_beam_heatmap}
% \end{figure}

\begin{figure}
\centering
\begin{minipage}[b]{1.0\linewidth}
\centerline{\includegraphics[width=\linewidth]{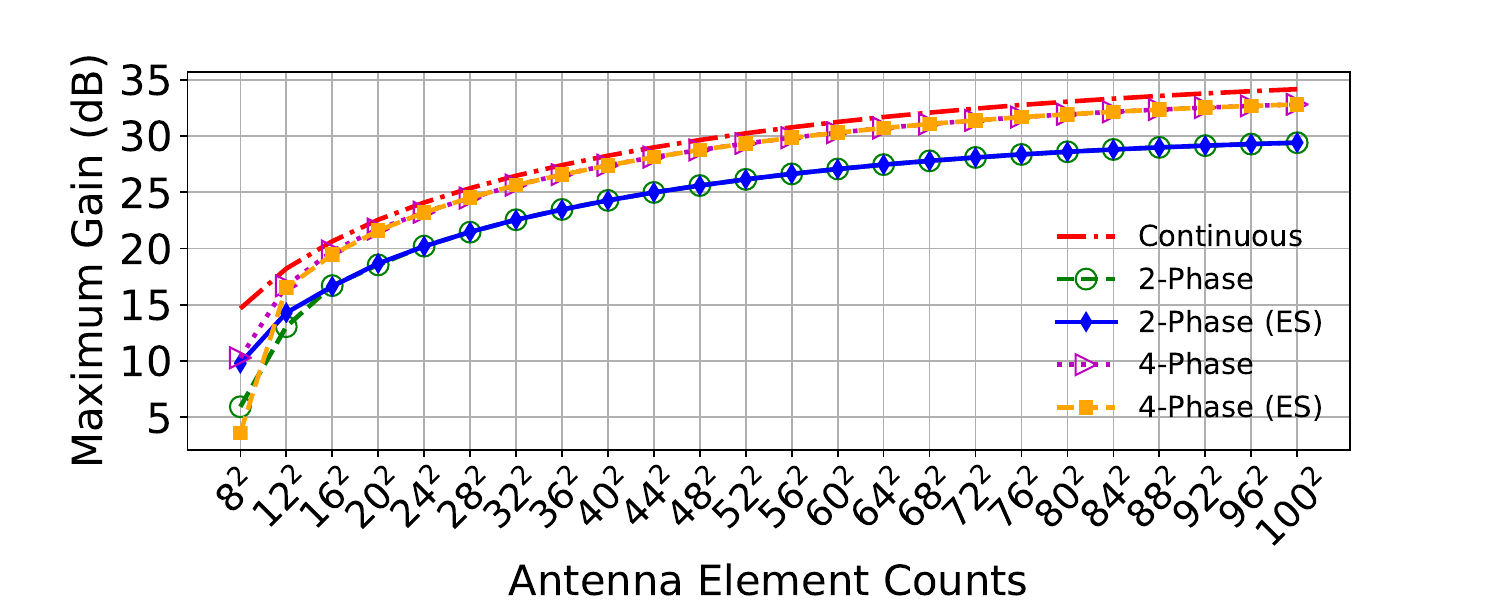}}
\centerline{\small (a) Maximum beam gain (dB).}
\label{f:gain_across_elements}
\end{minipage}

\begin{minipage}[b]{1.0\linewidth}
\centerline{\includegraphics[width=0.9\linewidth]{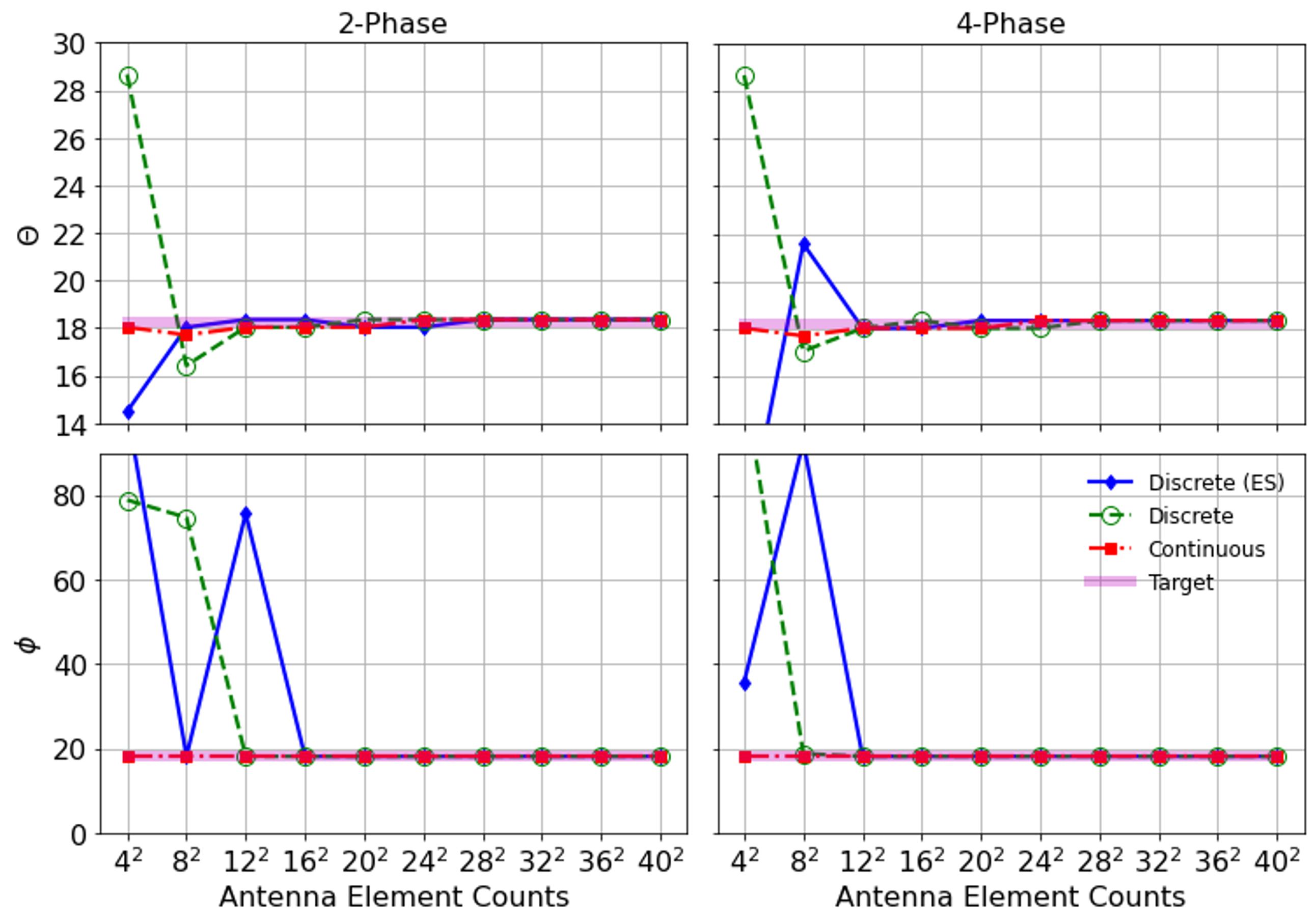}}
\centerline{\small (b) Beam direction with maximum gain.}
\label{f:direction_target_max}
\end{minipage}

\caption{Beamforming gains and directions across antenna element counts. ES in the figure stands for \emph{early stop} (Sec~\ref{s:early_stop_design}). 
% Both target $\phi$ and $\theta$ are 18.247 degrees.
} 
\label{f:beam_eval_comparison}
\end{figure}

% \begin{figure}
% \centering
% \includegraphics[width=0.9\linewidth]{example-image.png}
% \caption{Design, Series of QUBO} 
% % \label{f:cdf_ising}
% \end{figure}

% \begin{figure}
% \centering
% \includegraphics[width=\linewidth]{Figure/SA.png}
% \caption{Design, Series of QUBO} 
% % \label{f:cdf_ising}
% \end{figure}

% \begin{figure}
% \centering
% \includegraphics[width=0.95\linewidth]{Figure/stopping_criteria.png}
% \caption{Block Diagram of Stopping Criteria (DRAFT: Will be Updated)} 
% % \label{f:cdf_ising}
% \end{figure}

% \begin{figure}
% \centering
% \includegraphics[width=\linewidth]{Figure/antenna_gains_1.pdf}
% \caption{Block Diagram of Stopping Criteria (DRAFT: Will be Updated)} 
% % \label{f:cdf_ising}
% \end{figure}

% \begin{figure}
% \centering
% \includegraphics[width=\linewidth]{Figure/direction_data.png}
% \caption{Block Diagram of Stopping Criteria (DRAFT: Will be Updated)} 
% % \label{f:cdf_ising}
% \end{figure}

% \begin{figure}
% \centering
% \includegraphics[width=\linewidth]{Figure/beam_patterns.png}
% \caption{Block Diagram of Stopping Criteria (DRAFT: Will be Updated)} 
% % \label{f:b}
% \end{figure}

\section{Numerical Experiments}
\label{s:evaluation}

In this section, we evaluate our method with various beamforming scenarios. The radiating elements are point sources with omnidirectional patterns, and polarization is ignored. We assume no interaction or coupling between elements. Throughout the section, we denote our approach (\emph{i.e.,} discrete phase-shifting optimization based on a series of QUBO solved by the SA solver in the context of the bisection method) as discrete 2- or 4-phase shifting. The proposed early stop scheme (w/ \textbf{C1} and \textbf{C2}) is applied unless otherwise specified.
We consider square planar array geometries in our experiments.

\subsection{Benchmarking}
\label{s:evaluation_bechmark}

We first compare 3D beam patterns formed by different phase-shifting approaches in Figure~\ref{f:3D_benchmark}, where the arbitrary beam target direction is with $\theta$=$12.38^{\circ}, \phi$=$306.16^{\circ}$. We evaluate the ideal continuous weight (phase and amplitude) shifting (\emph{left}) and discrete weight 4-phase shifting with our solution approach (\emph{middle}). As an additional comparison scheme, we test a quantized version of the continuous weight (\emph{right}). For this, we quantize the optimal continuous weight that has an amplitude and phase into the closest available discrete phase (\emph{e.g.,} one of four possible phases), ignoring the amplitude information. Since both getting the optimal continuous weight and quantizing it into available discrete phases is simple, the method features a fast approximate solution.
Figure~\ref{f:3D_benchmark}(a) plots the beams for $16 \times 16$ array. Unlike the continuous phase shifting, unwanted sidelobes are observed in both 4-phase cases. However, the obtained beam patterns are quite similar given the small sidelobes, implying that even the simple quantized method can work efficiently for this array size. 
% Considering the search space of the phase shifting optimization problem (\emph{i.e.,} $4^{256}$), this is a surprising result. 
However, as we increase the array size to $32 \times 32$ 
% their beam patterns are 
, in the case of the quantized one, large gains of many unwanted sidelobes are observed as shown in Figure~\ref{f:3D_benchmark}(b), while the discrete phase shifting with our method can achieve similar beams to the optimal continuous setting in both $16 \times 16$ and $32 \times 32$ array scenarios (the direct quantitative comparison of the beam gains will be discussed in Sec~\ref{s:eval_beam}).
% with our approach has a similar beam pattern compared to $16 \times 16$ array but with stronger gains toward the target direction.
% Note that . 
To further analyze the beams, Figure~\ref{f:3D_benchmark}(c) plots normalized beam gain heatmaps. 
For $16 \times 16$ array scenarios, all of the schemes are able to focus the beam on the target area. However, for the $32 \times 32$ array scenarios, the quantized continuous one has two strong sidelobes near but out of the target area, while the discrete one still performs well.  
However, for large array sizes, it is sometimes observed that the discrete-phase shifting-based beamforming of our approach is slightly misaligned compared to the center of the target area for some directions, which does not occur with the optimal continuous setting.

To capture the difference between the 4-phase discrete shifting solved by our approach versus the quantized 4-phase one from the continuous phase, we plot the configuration of the continuous phase, discrete phase, and quantized discrete phase shifting for $8\times 8$ array in Figure~\ref{f:phase_configuration}. We observe these two discrete phase configurations are very different. While we show only the $8\times 8$ array scenario, it is the case for larger array sizes as well. This leads to two following conclusions. First, the solution of our optimization method cannot be estimated through the optimal continuous setting. Second, for relatively small arrays, approximate solutions that are attained from the optimal continuous setting are good enough even though the solutions are different from the optimal discrete configuration. However, as the size increases, this simple approximate method suffers from severe performance degradation (\emph{i.e.,} not scalable), while our approach achieves a similar beam pattern to the optimal continuous one. For this reason, we focus on the comparison between the optimal continuous one and our discrete approach without reporting the quantized one in the next subsection.

\subsection{Beamforming Evaluation}
\label{s:eval_beam}

We now quantitatively evaluate our method, comparing it against the optimal continuous weight case.

\begin{figure}
\centering
\includegraphics[width=0.9\linewidth]{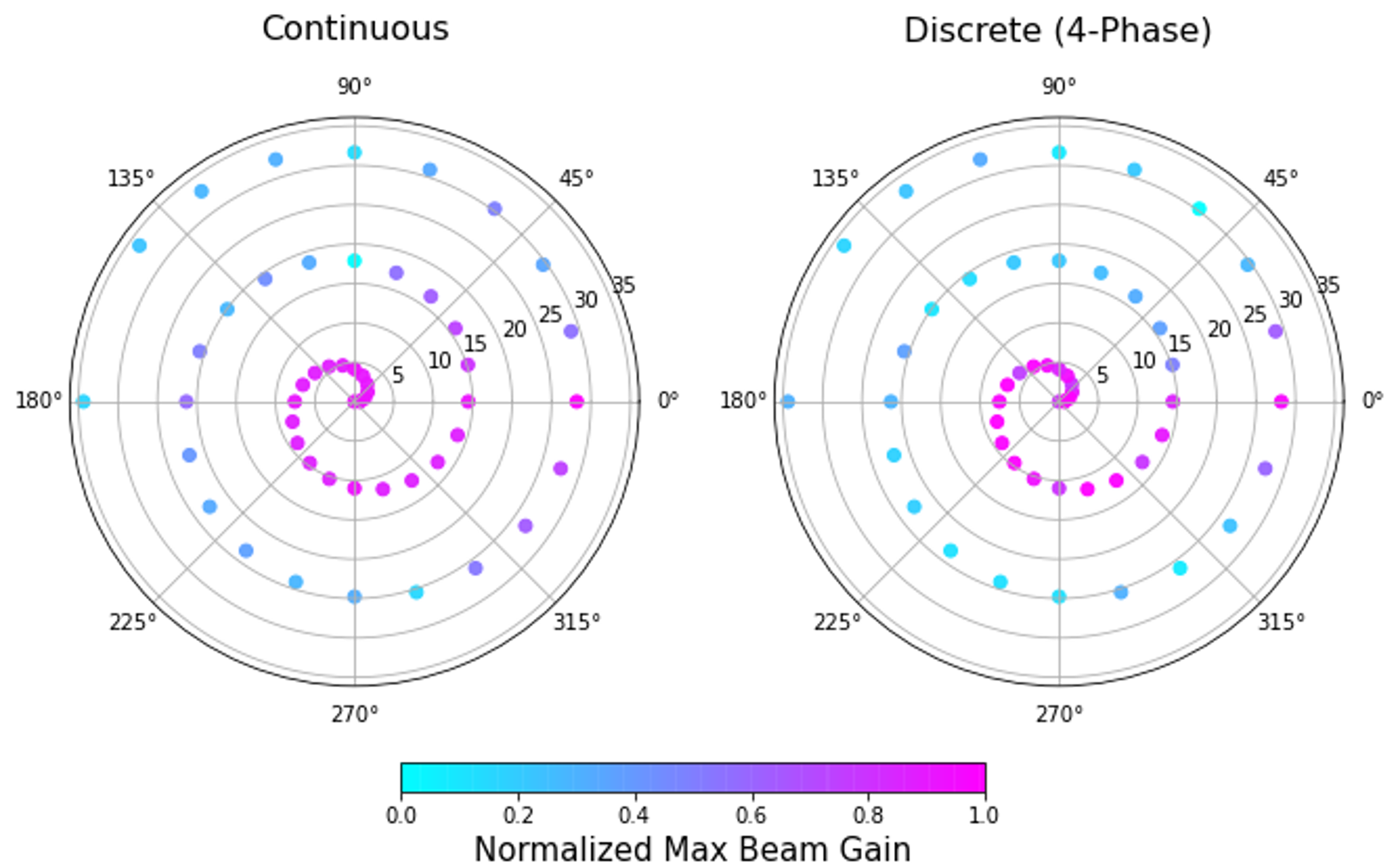}
\caption{Normalized beam gains across different target directions for $32\times32$ array scenarios with a spiral-shape exploration. The polar coordinate represents $\phi$, while the amplitude represents $\theta$. Gains are normalized separately.} 
\label{f:spiral}
\end{figure}

Figure~\ref{f:beam_eval_comparison}(a) shows the beamforming performance where both target $\phi$ and $\theta$ are 18.247 degrees. We plot the optimal continuous method and our discrete 2- and 4-phase shifting method with and without the proposed early stop (ES) scheme (Sec~\ref{s:early_stop_design}). When antenna element counts are small (\emph{e.g.,} $8^2$), discrete methods have poor and rather random performance, while the continuous setting can steer the beam precisely toward the target even with small antenna element counts. As the antenna element count increases, the performance of the discrete phase-shifting configuration becomes more stable, showing nearly constant gaps between their maximum gains and the maximum gains obtained by the optimal continuous setting. No noticeable performance gaps are observed for our discrete method with and without the early stop, which implies the proposed early stop method is an efficient strategy to reduce its computing time without loss of performance (achieved compute time gains will be discussed in the next subsection). Note that the discrete approach with 4 phases is able to achieve quite similar performance (less than \mbox{1--2~dB} gaps) to the ideal continuous setting, which is a similar phenomenon also observed in prior work~\cite{optimalbeamforming2022} in the context of the RIS scenarios. Compared to the 4-phase setting, the gains of the 2-phase setting are lower gains. However, given that the gaps between the 2-phase case and the optimal continuous case are still insignificant (less than 5~dB), we still believe
the 2-phase discrete configuration is also a good candidate approach toward large-scale antenna array for beamforming, especially considering the simplicity of 2-phase hardware implementation (it is much simpler than even the 4-phase case). Figure~\ref{f:beam_eval_comparison}(b) presents the beamforming directions where the maximum beam gains are formed. As the array size increases, the discrete method with both 2-phase and 4-phase can precisely steer the beam toward a target direction. This implies the maximum gains we saw in Figure~\ref{f:beam_eval_comparison}(a) are well formed in the target beam direction.

Figure~\ref{f:spiral} shows maximum beam gains for more various target directions through a spiral-shape exploration: (\emph{left}) continuous and (\emph{right}) discrete 4-phase setting. Beam gains are normalized by the achieved maximum gain (at $\theta=0^{\circ}$, $\phi=0^{\circ}$). In both 2-phase and 4-phase cases, we found that beam steering is more vulnerable to the elevation ($\theta$) than to the azimuth ($\phi$). Compared to the continuous one, the discrete one features relatively limited elevation steering ability, showing relatively weak beams from around $\theta=15^{\circ}$. However, even the continuous setting starts suffering from performance degradation near $\theta=20^{\circ}$. Furthermore, the distance between the transmitter and receiver in commonly used far-field scenarios could make this elevation issue minor. Considering the observed beamforming ability that works for any $\phi$ (as long as with $\theta \leqslant 15^{\circ}$), we believe the proposed design demonstrates a promising beamforming ability.

\begin{figure}
\centering
\includegraphics[width=\linewidth]{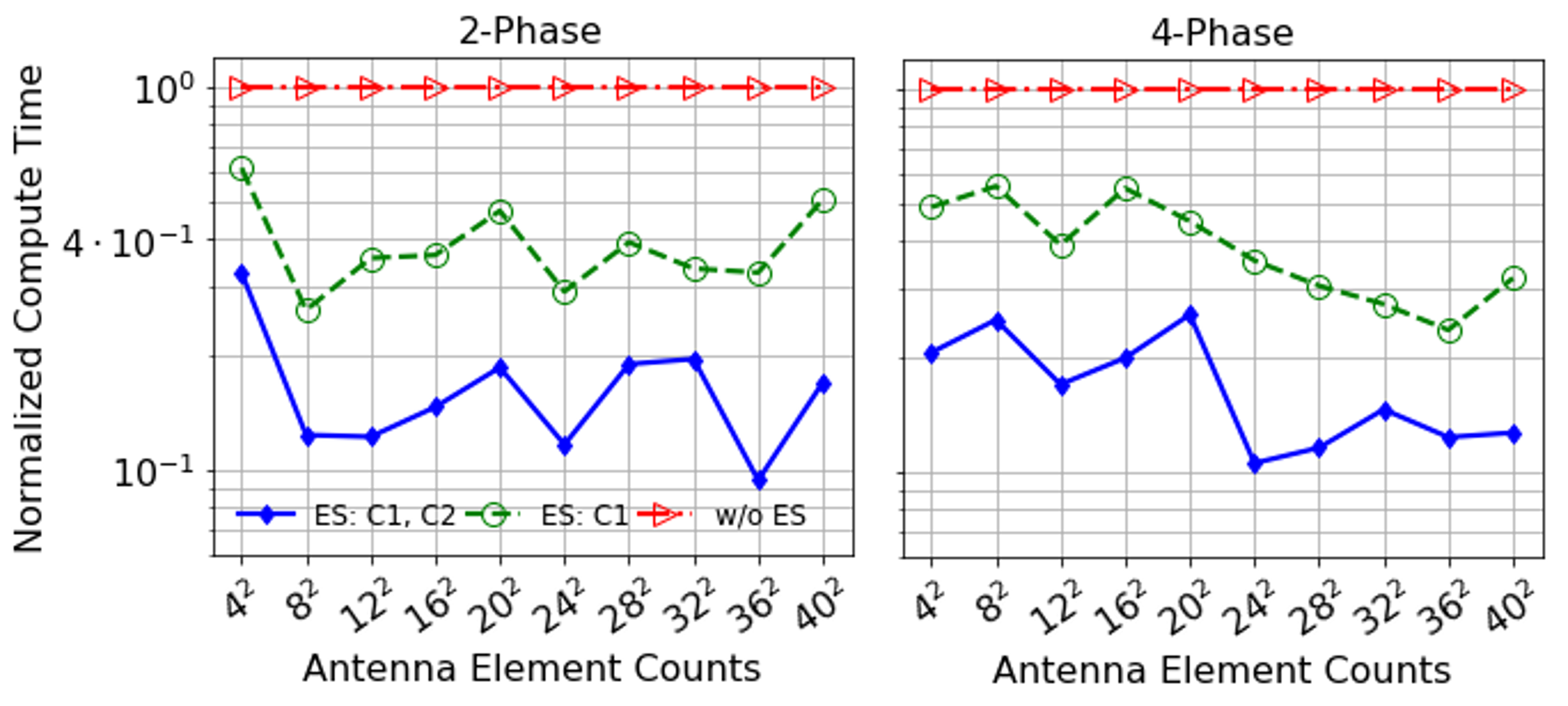}\vspace{-0.15cm}
\caption{Normalized compute time across antenna element counts for SA with different stopping criteria applied.}\vspace{-0.15cm} 
\label{f:compute_time}
\end{figure}

% we believe this is quite a promising beamforming result.
% However, this relativeness comes from the strong max beam at $\theta=0^{\circ}$, $\phi=0^{\circ}$ with large arrays, 

\subsection{Compute Time Evaluation}
\label{s:eval_compute_time}

\begin{figure}
\centering
\includegraphics[width=\linewidth]{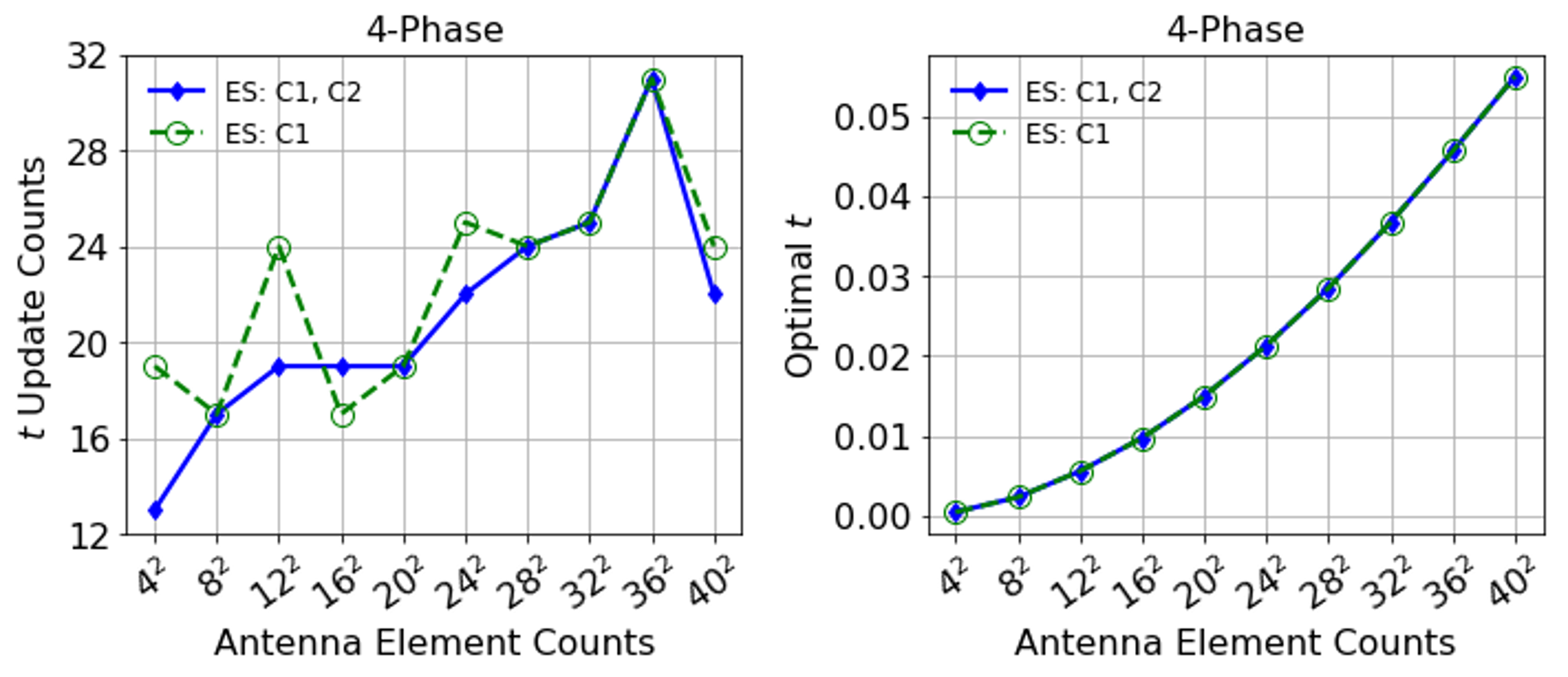}\vspace{-0.15cm}
\caption{Number of $t$ updates (equivalent to the number of the QUBO forms solved in the series for the bisection method) and optimal $t$ across antenna element counts.}\vspace{-0.15cm} 
\label{f:t_update}
\end{figure}

To evaluate the proposed stopping criteria and early stop scheme further, we show the normalized compute time for our method with and without the early stop scheme in Figure~\ref{f:compute_time}, where compute time is represented as the number of bits explored during the entire SA processing. In the case of the method without the early stop scheme, it features a predictable compute time since it explores all the bits during 50 sweeps and 50 batches (default setting). When with the early stop scheme, up to 10$\times$ speedup is achieved in both 2- and 4-phase cases. Interestingly, we observe that applying \textbf{C1} only can reduce the compute time by around 60\% compared to the one without the scheme, and applying \textbf{C2} on top of \textbf{C1} is able to reduce the compute time by nearly 50\% further. 
Since we observed that the method with the early stop scheme does not suffer from performance degradation in the previous subsection (Sec~\ref{s:eval_beam}), we can argue that applying our proposed stopping criteria for early stops to the solver is an efficient way to accelerate the SA process without critical loss of optimization and thus beamforming performance. We also demonstrate this by showing that nearly the same final $t$ values are obtained with and without the early stop, despite different QUBO counts solved in the process as shown in Figure~\ref{f:t_update}. 
% Recall that the result of one with only \textbf{C1} stopping criterion is the same as the one without ES. 

% maximum beam gains for different phase shifting schemes (continuous shifting, discrete 2- and 4-phase shifting) in Figure~\ref{f:beam_comparison} (a). The beam target direction is  
% \subsection{Numerical Experiments}
% \label{s:evaluation}

% \begin{table}[t]
% \caption{notation table}
% \begin{tabularx}
% {\linewidth}{*{6}{X}}
% \toprule
%   \multicolumn{2}{l}{{\ul Indices:}}   \\
%   $K$ & number of products ($k \in\{1,...,K\}$)  \\ 
%   $T$ & number of periods ($t \in\{1,...,T\}$)\\
%   \multicolumn{2}{l}{{\ul Parameters:}}  \\  
%   $\delta_k$ & target δ-service level for product k \\
%   $\epsilon$ &  \\
%   $c_t$      & production capacity in period t \\
%   $c^r_t$    & Remanufacturing capacity in period t\\
%   $pc_{k}$   & production cost of product k per unit\\
%   $pc^r_{k}$ & remanufacturing cost of product k per unit\\
% \bottomrule
% \end{tabularx}
% \end{table}

\section{Conclusion}
\label{s:conclusion}
In this work, physics-inspired discrete phase shifting optimization for 3D beamforming with large-scale PIN-diode antenna arrays has been investigated. To leverage the physics-inspired methods, the beamforming optimization problem has been translated into an equivalent QUBO structure consisting of sequential QUBO forms representing the bisection method, which is a very distinctive QUBO formulation compared to other work. As an initial guiding framework, we use an SA-based method as a QUBO solver in the structure, showing that the resulting
discrete phase configuration
with 4-phase shifting can achieve a similar beamforming
performance to one that the optimal continuous setting
can achieve even for extremely large 10,000-element scenarios. 
% This scale is 2500$\times$ larger than conventional 
% In the case of 2-phase shifting, while it has slightly lower maximum beam gains compared to the optimal setting, its gains still increase gently as the size of the antenna array increases. Given that the gaps are less than 5~dB between the optimal continuous setting and 2-phase setting and the simplicity of 2-phase hardware implementation, we believe the 2-phase discrete configuration is also a good candidate approach toward large-scale antenna array for beamforming.
% Our QUBO formulation is unique in that the optimization problem of interest is expressed as a series of QUBO  
% Considering the unique objective of each QUBO in the sequential series structure in our formulation, we design stopping criteria for early stops. 
% this difference, we adapt the simulated annealing algorithm with stopping criteria. 
% We show that discrete 4-phase shifting configuration with our method can achieve quite similar beamforming performance to the optimal continuous setting. 
% It is observed that our proposed early stop scheme with stopping criteria in SA can reduce overall compute time by nearly $10\times$ without loss of beam gains on the target. 
% We adopt an SA algorithm and apply an early stop with stopping criteria, considering the unique objective of each QUBO in the structure. 

Note that SA is a rudimentary physics-inspired optimization method that has led to many advanced algorithms and emerging computing devices.
% such as quantum annealers, digital annealers, coherent Ising machines (CIM), and optical Ising machines (OIM). 
Given the unique objective of each QUBO in our formulation, tweaking such algorithms and/or hardware implementations can be a compelling research topic to explore. 
In this regard, it is worth mentioning that the quantum approximate optimization algorithm (QAOA)~\cite{farhi2014quantum} that can be conducted on the gate model quantum computers is desirable for the structure in that, unlike QA, it requires extra steps between iterations (\emph{i.e.,} controllability between iterations). While available qubit counts on the current gate-model quantum computers are too small to conduct large-scale array beamforming experiments, in light of quantum devices with large-scale qubits, we believe a physics-inspired approach with gate-model quantum hardware is worth exploring for this research direction.

\section*{\large Acknowledgement}
This research is based upon work supported by the National Science Foundation under Grant No. CNS-1824357.

% Further  future work, we will investigate the implementation of the proposed early stop scheme for fully parallel SA processing, where all the batches are executed in parallel.  like parallel tempering Monte Carlo and

% \section*{Acknowledgments}
% This should be a simple paragraph before the References to thank those individuals and institutions who have supported your work on this article.


% \section{References Section}
% You can use a bibliography generated by BibTeX as a .bbl file.
%  BibTeX documentation can be easily obtained at:
%  http://mirror.ctan.org/biblio/bibtex/contrib/doc/
%  The IEEEtran BibTeX style support page is:
%  http://www.michaelshell.org/tex/ieeetran/bibtex/
 
%  % argument is your BibTeX string definitions and bibliography database(s)
% %\bibliography{IEEEabrv,../bib/paper}
% You can manually copy in the resultant .bbl file and set second argument of $\backslash${\tt{begin}} to the number of references
%  (used to reserve space for the reference number labels box).

\clearpage

\begin{raggedright}
\bibliographystyle{concise2}
\bibliography{reference}
\end{raggedright}

% \newpage

 

% \end{IEEEbiography}

\vspace{11pt}


% \vfill

\end{document}